\documentclass[a4paper,11pt,preprintnumbers]{article}
\pdfoutput=1 

\usepackage{jheppub} 
\usepackage{amsfonts}
\usepackage{amssymb,amsmath}
\usepackage{hyperref}
\usepackage{comment}
\usepackage{graphicx}
\usepackage{subcaption}
\usepackage{color,graphicx,slashed,multirow,booktabs,multirow}
\usepackage{braket}
\usepackage{bbold}
\usepackage{xcolor}
\usepackage[utf8]{inputenc}
\usepackage{subfloat}
\usepackage{cleveref}
\usepackage{ulem}
\usepackage{eqparbox}

\usepackage{mathtools}

\usepackage[force]{feynmp-auto}
\usepackage[T1]{fontenc} 

\newcommand{\lm}[1]{\textcolor{violet}{#1}}

\newcommand{\equaref}[1]{Eq.~(\ref{#1})}

\newcommand{\appref}[1]{Appendix~\ref{#1}}

\newcommand{\mybox}[1]{\boxed{\phantom{{}^a_1}\!\! #1}}
\newcommand{\myybox}[1]{\boxed{\phantom{{}^a_j}\!\! #1}}

\title{Non-Abelian Domain Walls and Gravitational Waves}
\preprint{IPPP/24/62}

\author[a,h]{Bowen Fu,}
\affiliation[a]{Key Laboratory of Cosmology and Astrophysics (Liaoning) \& College of Sciences, Northeastern University, Shenyang 110819, China}
\author[b]{Stephen F. King,}
\affiliation[b]{School of Physics and Astronomy, University of Southampton, Southampton, SO17 1BJ, U.K.}
\author[c]{Luca Marsili,}
\affiliation[c]{Instituto de F\`isica Corpuscular (IFIC), Universitat de Val\`encia, Parc Cientific UV, C/ Catedratico Jose Beltran 2, E-46980 Paterna, Spain}
\author[d,e]{Silvia Pascoli,}
\affiliation[d]{DIFA, University of Bologna, via Irnerio 46, 40126 Bologna, Italy}
\affiliation[e]{INFN, Sezione di Bologna, viale Berti Pichat 6/2, 40127 Bologna, Italy}
\author[f]{Jessica Turner}
\affiliation[f]{Institute for Particle Physics Phenomenology, Department of Physics, Durham University, Durham DH1 3LE, U.K.}
\author[g]{and Ye-Ling Zhou}
\affiliation[g]{School of Fundamental Physics and Mathematical Sciences, Hangzhou Institute for Advanced
Study, UCAS, Hangzhou 310024, China}
\affiliation[h]{Foshan Graduate School of Innovation, Northeastern University, Foshan 528312, China}

\emailAdd{fubowen@mail.neu.edu.cn}
\emailAdd{s.f.king@soton.ac.uk}
\emailAdd{luca.marsili@ific.uv.es}
\emailAdd{silvia.pascoli@unibo.it}
\emailAdd{jessica.turner@durham.ac.uk}
\emailAdd{zhouyeling@ucas.ac.cn}

\abstract{
We investigate the properties of domain walls arising from non-Abelian discrete symmetries, which we refer to as non-Abelian domain walls. We focus on $S_4$, one of the most commonly used groups in lepton flavour mixing models.
The spontaneous breaking of $S_4$ leads to distinct vacua preserving a residual $Z_2$ or $Z_3$ symmetry. 
Five types of domain walls are found, labelled as SI, SII, TI, TII, and TIII, respectively, the former two separating $Z_2$ vacua and the latter three separating $Z_3$ vacua. We highlight that SI, TI and TIII may be unstable for some regions of the parameter space and decay to stable domain walls. Stable domain walls can collapse and release gravitational radiation for a suitable size of explicit symmetry breaking. A symmetry-breaking scale of order 100 TeV may explain the recent discovery of nanohertz gravitational waves by PTA experiments. For the first time, we investigate the properties of these domain walls, which we obtain numerically with semi-analytical formulas applied to compute the tension and thickness across a wide range of parameter space. We estimate the resulting gravitational wave spectrum and find that, thanks to their rich vacuum structure, non-Abelian domain walls manifest in a very interesting and complex phenomenology. 
 }

\keywords{non-Abelian flavour symmetry, domain wall, gravitational waves}

\begin{document}

\maketitle


\section{Introduction}
Since the discovery of neutrino oscillations \cite{Kamiokande-II:1990wrs,Cleveland:1998nv,GALLEX:1998kcz,SAGE:1999nng,Super-Kamiokande:2001ljr}, the origin of neutrino masses and leptonic mixing remains the most concrete evidence of physics beyond the Standard Model. 
Compared with the CKM quark mixing, leptonic mixing is significantly larger.
To explain the leptonic mixing pattern, the most studied approach is that of flavour symmetries (or family symmetries), among which non-Abelian discrete symmetries are the most popular candidate \cite{Altarelli:2010gt,King:2013eh,King:2017guk,Xing:2020ijf}.
The scale at which these flavour symmetries break is typically assumed to be very high making them largely inaccessible at current collider experiments.

Since their discovery, gravitational waves (GW) have become a powerful probe of new physics at high scales \cite{LIGOScientific:2016aoc}. Moreover, recently there has been the discovery of a stochastic background of nanohertz gravitational waves by pulsar timing array experiments \cite{NANOGrav:2023gor,EPTA:2023fyk,EPTA:2023xxk,Reardon:2023gzh,Xu:2023wog} which may be of a cosmological origin \cite{Madge:2023dxc}. Among the cosmological sources of gravitational waves, topological defects \cite{Vilenkin:1984ib,Kibble:1976sj,Preskill:1984gd} including cosmic strings \cite{Auclair:2019wcv,Gouttenoire:2019kij}, domain walls \cite{Hiramatsu:2010yz,Kawasaki:2011vv,Hiramatsu:2013qaa,Saikawa:2017hiv}, and hybrid defects \cite{Buchmuller:2021mbb,Dunsky:2021tih}, provide a strong connection between high-scale new physics and the gravitational wave observations \cite{King:2020hyd,King:2021gmj,Fu:2022lrn,Fu:2023mdu}. 
It has been highlighted that GWs from collapsing domain walls provide a good fit to the data of the recent NANOGrav Pulsar Time Array (PTA) measurement \cite{NANOGrav:2023hvm}. Several studies have explored this signal, with most focusing on $Z_2$ domain walls or axion domain walls \cite{Kitajima:2023cek,Bai:2023cqj, Babichev:2023pbf,Zhang:2023nrs, Blasi:2023sej, Gouttenoire:2023ftk, Ferreira:2024eru}. General discussions on the classification of domain walls from the cyclic group $Z_N$ (for $N>2$) and the consequent GW signals are presented in \cite{Wu:2022tpe,Wu:2022stu}.
It was pointed out that non-Abelian discrete symmetries can also lead to domain walls and thus observable gravitational waves \cite{Gelmini:2020bqg}. However, the analysis of domain walls of Ref.~\cite{Gelmini:2020bqg} was qualitative as it was based on the results for the $Z_2$ symmetry. Different configurations of domain walls generated from non-Abelian discrete symmetries have also been discussed in \cite{Jueid:2023cgp, Yang:2024bys}. The domain wall properties from non-Abelian discrete symmetry breaking and the resulting GW signal are still unexplored.

In this work, we perform a detailed analysis of the domain walls from non-Abelian discrete symmetry breaking. 
We consider the group to be the octahedral group $S_4$, which is a subgroup of $SO(3)$ and one of the most popular groups used in addressing the origin of lepton flavour mixing as its predictions fit data well \cite{Hagedorn:2012ut,Ding:2013eca,Luhn:2013lkn,King:2016yvg}. In the present work, we will focus on one triplet flavon of $S_4$ to simplify the discussion.
By analysing the vacuum structure of a general renormalisable scalar potential respecting this symmetry, we find that there are two different types of vacua, one preserving $Z_2$ and the other preserving $Z_3$ symmetries.
For each type of vacuum, there are different types of domain walls (five types in total) that depend on the topology of the vacuum manifold.
We point out that some domain walls are unstable in part of the parameter space, as their energies are high enough to enable them to decay to lower-energy domain walls. 
We also estimate the gravitational wave produced by stable non-Abelian domain walls, in the presence of suitable explicit symmetry breaking. 
Due to different types of domain walls, the gravitational wave spectrum has a richer structure than the simple $Z_2$ case.

The paper is organised as follows. We take $S_4$ as a representative example and discuss the vacuum structure based on the most generic renormalisable potential in Sec.~\ref{sec:vac}. 
Then we present the domain wall solutions for each type of vacuum in Sec.~\ref{sec:DW}. 
In Sec.~\ref{sec:stability}, we clarify the stability of the domain wall solutions obtained in Sec.~\ref{sec:DW}. 
Finally, we estimate the gravitational wave signal produced by the non-Abelian domain walls in Sec.~\ref{sec:GW} and summarise in Sec.~\ref{sec:con}. 

\section{Flavon potential and vacuum alignment \label{sec:vac} }
We begin our discussion with a brief introduction to the octahedral group $S_4$. 
$S_4$ is the group of permutations of four objects. It contains 24 elements, with three generators $S$, $T$ and $U$ satisfying $S^2=T^3=(ST)^3=U^2=(SU)^2=(TU)^2=(STU)^4=1$. 
There are five irreducible representations (irrep), the trivial one-dimensional (1d) irreps ${\bf 1}$, another 1d sign representation ${\bf 1}'$, one two-dimensional (2d) irrep ${\bf 2}$, and two three-dimensional (3d) irreps ${\bf 3}$ and ${\bf 3}'$. 
The $S$, $T$ and $U$ generators of $S_4$ in 3d irreps are given by
\begin{eqnarray}
T=\left(
\begin{array}{ccc}
 0 & 0 & 1 \\
 1 & 0 & 0 \\
 0 & 1 & 0 \\
\end{array}
\right)\,,\quad
S=\left(
\begin{array}{ccc}
 1 & 0 & 0 \\
 0 & -1 & 0 \\
 0 & 0 & -1 \\
\end{array}
\right)\,,\quad
U = \pm \begin{pmatrix}
 1 & 0 & 0 \\ 0 & 0 & 1 \\ 0 & 1 & 0
 \end{pmatrix}\,,
\label{eq:generator1}
\end{eqnarray} 
where the $\pm$ sign of $U$ is for ${\bf 3}$ and ${\bf 3}'$, respectively. 
Here, we have applied the 3d representation in the Ma-Rajasekaran basis of $S$ and $T$ \cite{Ma:2001dn}. 
Compared with the Altarelli-Feruglio basis \cite{Altarelli:2005yx}, which is widely used in flavour model building, it is easier for us to find the vacua of $\phi$ in this basis and the physics remains equivalent.

We introduce a real flavon triplet $\phi=(\phi_1, \phi_2, \phi_3)^T$. 
By arranging the flavon as a ${\bf 3}'$ of $S_4$, the most general renormalisable flavon potential is
\begin{eqnarray} \label{eq:potential}
V(\phi) = -\frac{\mu^2}{2} I_1 + \frac{g_1}{4} I_1^2 + \frac{g_2}{2} I_2\,,
\end{eqnarray}
where 
\begin{eqnarray}
I_1 &=& \phi_1^2+\phi_2^2+\phi_3^2 \,, \nonumber\\
I_2 &=& \phi_1^2\phi_2^2+\phi_2^2\phi_3^2+\phi_3^2\phi_1^2 \,. 
\end{eqnarray}
To ensure the potential {positive}, $g_1>0$ and $g_2> - 4 g_1$ are required. 
We provide details of the potential construction in \appref{app:A}. 
Note that if $\phi$ is arranged as a ${\bf 3}$ of $S_4$, a cubic term $\phi_1 \phi_2 \phi_3$ is allowed in the potential. 
Its effect on the vacuum structure and domain wall properties can be found in \appref{app:C}.

A necessary condition for the vacuum of $\phi$ is $\partial V(\phi)/\partial \phi_i=0$.
The solutions satisfying this condition can be divided into three classes according to the corresponding values of $V(\phi)$: 
\begin{eqnarray}\label{eq:vacua}
\text{(1)}&&v_m \in \left\{\begin{pmatrix} 1 \\ 0 \\ 0 \end{pmatrix}, 
\begin{pmatrix} 0 \\ 1 \\ 0 \end{pmatrix},
\begin{pmatrix} 0 \\ 0 \\ 1 \end{pmatrix},
\begin{pmatrix} -1 \\ 0 \\ 0 \end{pmatrix}, 
\begin{pmatrix} 0 \\ -1 \\ 0 \end{pmatrix},
\begin{pmatrix} 0 \\ 0 \\ -1 \end{pmatrix}
\right\}v\,,
\hspace{12mm}\text{for}~ m=1,2,...6;\nonumber\\
\text{(2)}&&u_n=\left\{\begin{pmatrix} 1 \\ 1 \\ 1 \end{pmatrix}, 
\begin{pmatrix} -1 \\ 1 \\ 1 \end{pmatrix},
\begin{pmatrix} 1 \\ -1 \\ 1 \end{pmatrix},
\begin{pmatrix} 1 \\ 1 \\ -1 \end{pmatrix},
\begin{pmatrix} -1 \\ -1 \\ -1 \end{pmatrix}, 
\begin{pmatrix} 1 \\ -1 \\ -1 \end{pmatrix},
\begin{pmatrix} -1 \\ 1 \\ -1 \end{pmatrix},
\begin{pmatrix} -1 \\ -1 \\ 1 \end{pmatrix}
\right\}u\,,\nonumber\\
&&\hspace{106mm}
 \text{for}~ n=1,2,...8;
 \nonumber\\
\text{(3)}&&s_l \in \left\{\begin{pmatrix} 0 \\ 1 \\ 1 \end{pmatrix}, 
\begin{pmatrix} 1 \\ 0 \\ 1 \end{pmatrix},
\begin{pmatrix} 1 \\ 1 \\ 0 \end{pmatrix},
\begin{pmatrix} 0 \\ 1 \\ -1 \end{pmatrix},
\begin{pmatrix} -1 \\ 0 \\ 1 \end{pmatrix},
\begin{pmatrix} 1 \\ -1 \\ 0 \end{pmatrix},\right.\nonumber\\
&&\hspace{1cm} \left.
\begin{pmatrix} 0 \\ -1 \\ -1 \end{pmatrix},
\begin{pmatrix} -1 \\ 0 \\ -1 \end{pmatrix},
\begin{pmatrix} -1 \\ -1 \\ 0 \end{pmatrix},
\begin{pmatrix} 0 \\ -1 \\ 1 \end{pmatrix},
\begin{pmatrix} 1 \\ 0 \\ -1 \end{pmatrix},
\begin{pmatrix} -1 \\ 1 \\ 0 \end{pmatrix}\right\}s\,,\hspace{6mm}
 \text{for}~ l=1,2,...,12\,, \nonumber\\
\label{eq:solution1}
\end{eqnarray}
where
\begin{eqnarray}
&&v=\frac{\mu}{\sqrt{g_1}}\,, \hspace{2.5cm}
V(v_m)=-\frac{\mu^4}{4g_1}\,,\nonumber\\
&&u=\frac{\mu}{\sqrt{3g_1+2g_2}}\,, \hspace{1.3cm}
V(u_n)=-\frac{3\mu^4}{4 (3 g_1 + 2g_2)}\,,\nonumber\\
&&s=\frac{\mu}{\sqrt{2g_1+g_2}}\,, \hspace{1.5cm}
V(s_l)=-\frac{\mu^4}{4 g_1 + 2g_2}\,.
\end{eqnarray}
The first class of solutions, $v_m$, preserves a different $Z_2$ symmetry. For example, $(1,0,0)^T$ is invariant under a $Z_2$ transformation generated by $S$, and $(0,1,0)^T$ is invariant under a $Z_2$ transformation generated by $TST^2$. 
Similarly, each $u_n$ in the second class of solutions preserves a different $Z_3$ symmetry, e.g., $(1,1,1)^T$ invariant in a $Z_3$ generated by $T$ and $(-1,1,1)^T$ is invariant under a $Z_3$ transformation generated by $STS$.
For convenience, $v_m$ and $u_m$ will also be denoted as $Z_2$-preserving and $Z_3$-preserving vacua in the remainder of the paper, respectively. 

\begin{figure}[t!]
\centering
\includegraphics[width=.32\textwidth]{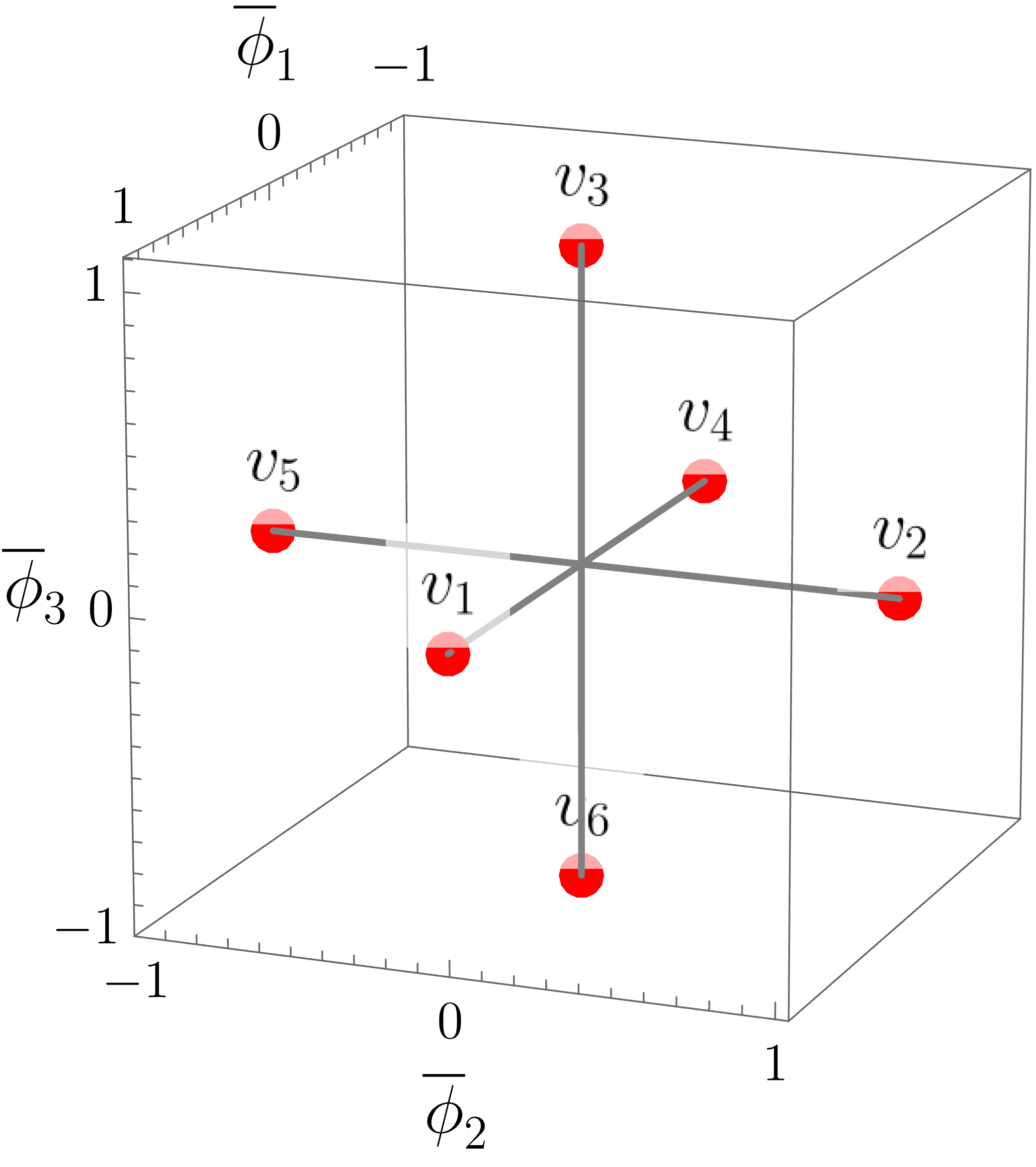} \hspace{2cm}
\includegraphics[width=.32\textwidth]{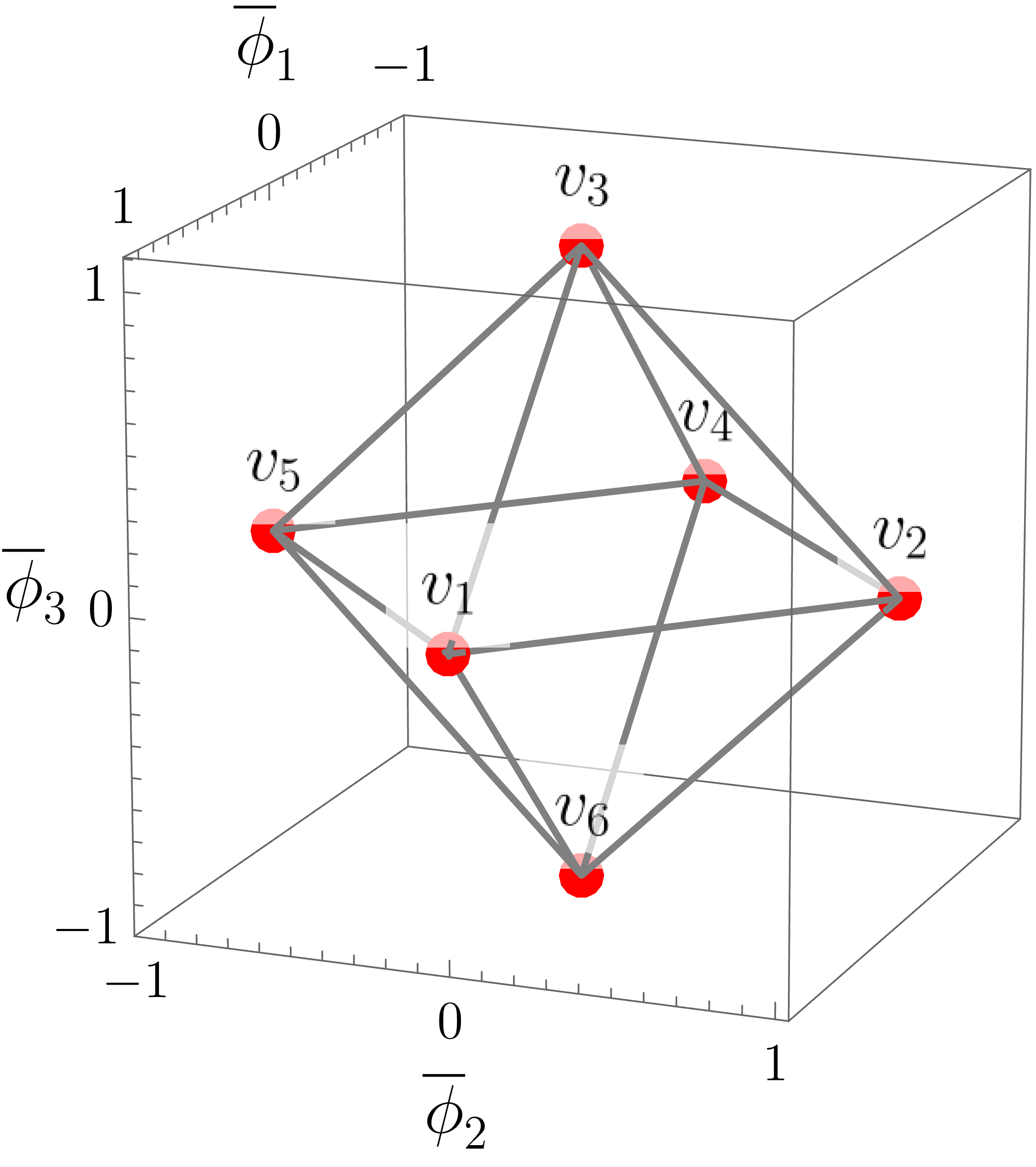}
\caption{$Z_2$-preserving vacua of $S_4$ in the three-dimensional flavon field space and two topologically different types of DWs. A straight line connecting to two vacua should not be understood as the path in the field space, but just represent the topology of the DW separating these two vacua. SI and SII DWs are given in the left and right panels, respectively.}
\label{fig:DW_S}
\end{figure}

\begin{figure}[t!]
\centering
\includegraphics[width=.32\textwidth]{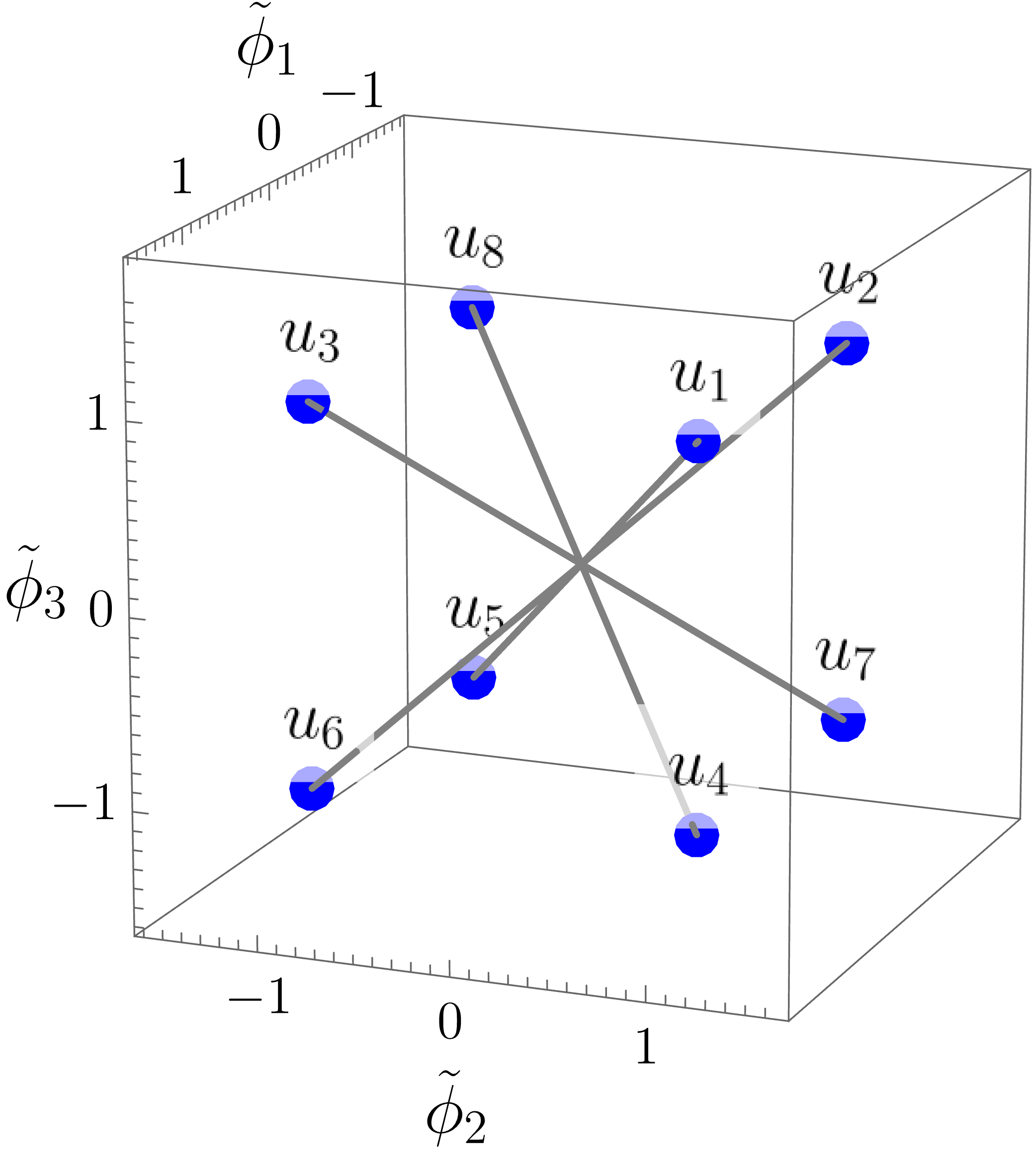}
\includegraphics[width=.32\textwidth]{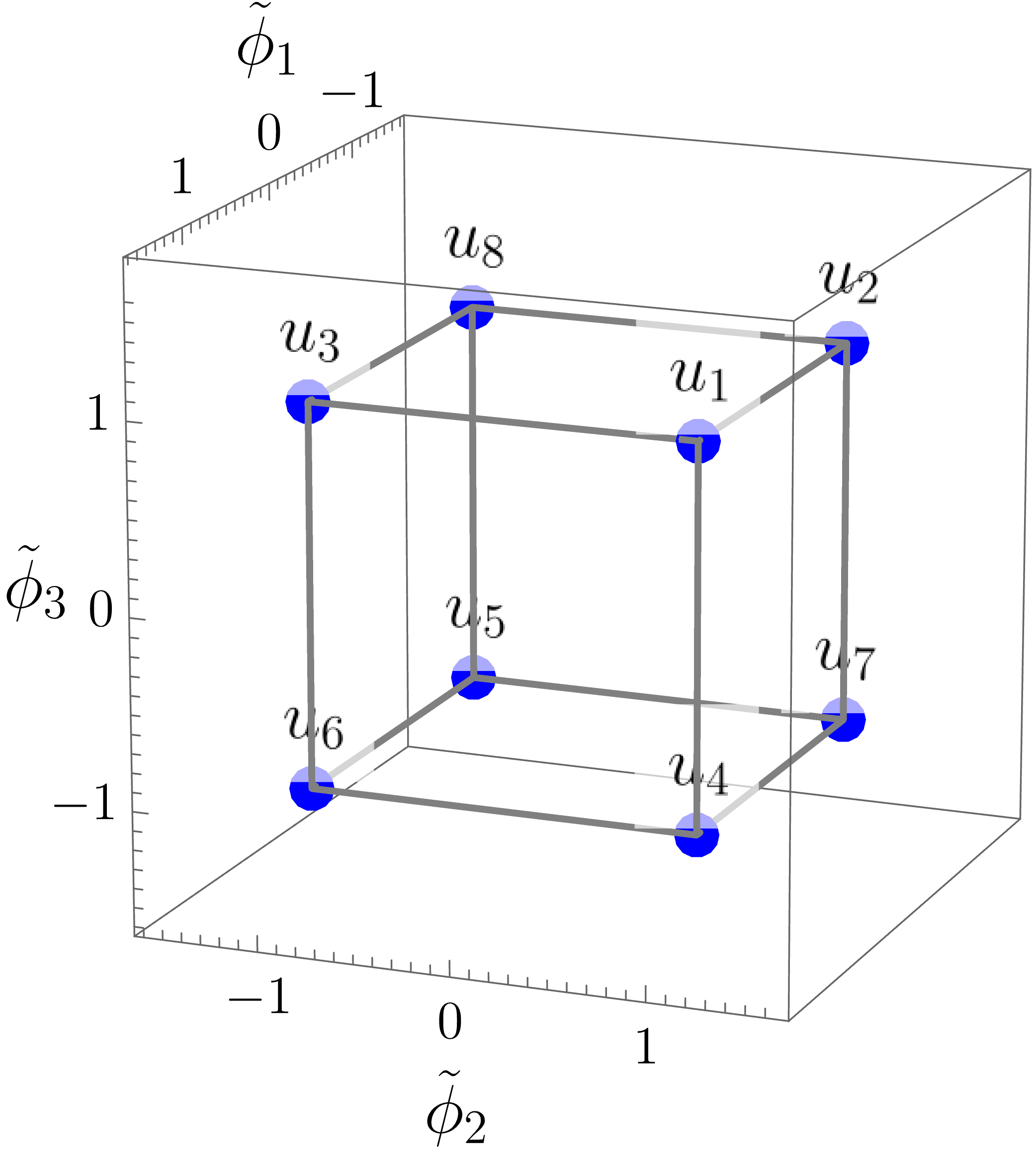}
\includegraphics[width=.32\textwidth]{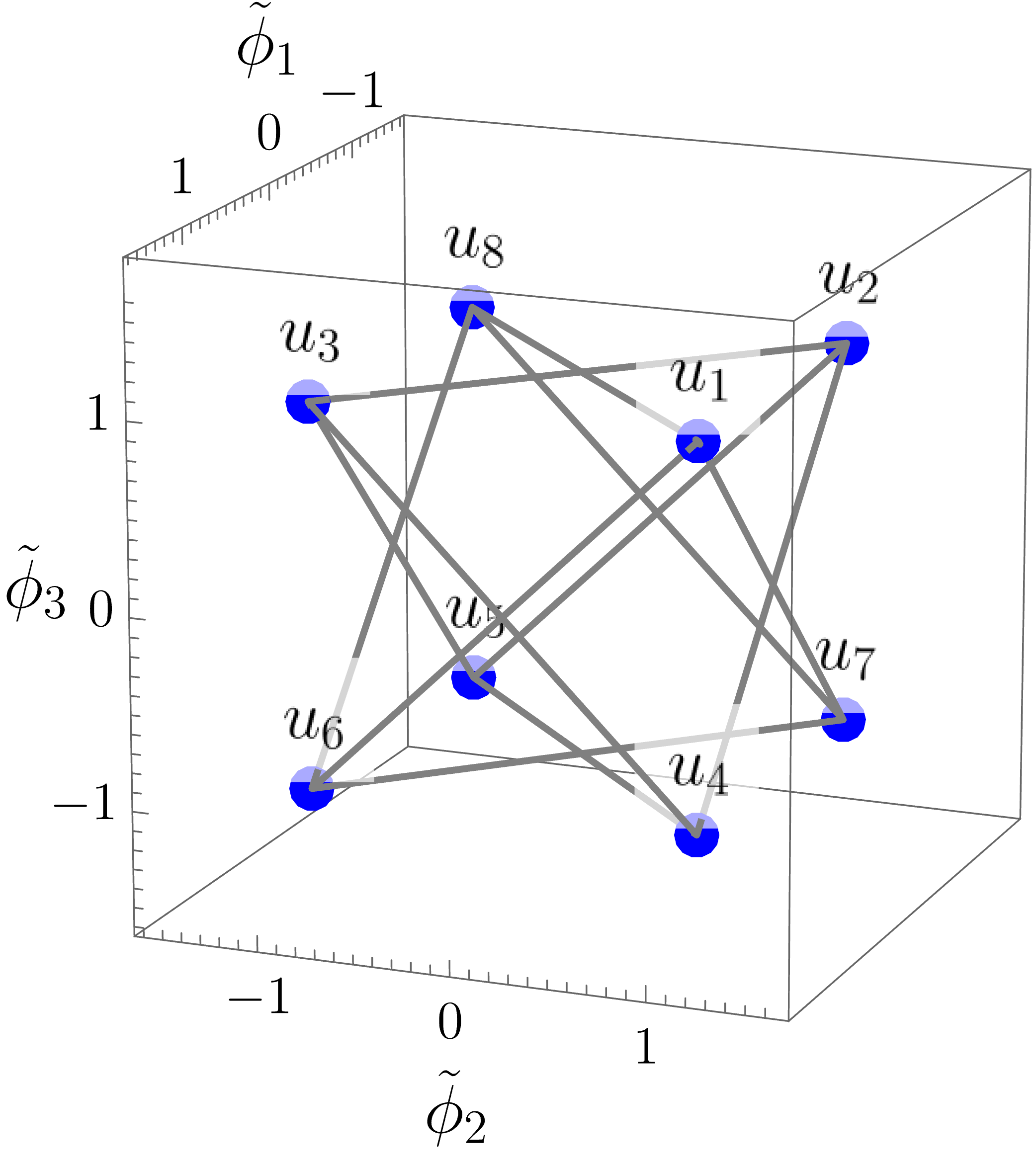}
\caption{$Z_3$-preserving vacua of $S_4$ and three topologically different types of DWs. TI, TII and TIII DWs are given in the left, middle and right panels.}
\label{fig:DW_T}
\end{figure}

To guarantee that some of the above solutions are vacua, $V(\phi)$ must be a local minimum at these solutions, corresponding to the requirement of a positive-definite second derivative of $V(\phi)$. 
More specifically, the matrix $M_\phi^2$ defined in the following should be positive-definite at these solutions: 
\begin{eqnarray}
(M_\phi^2)_{ij}=\frac{\partial^2 V(\phi)}{\partial \phi_i \partial \phi_j}\Big|_{\langle \phi \rangle} \,.
\end{eqnarray}
In general, $M_\phi^2$ is a $3\times 3$ real symmetric matrix which can be diagonalised through $W^T M_\phi^2 W=\text{diag}\{m_{1}^2,m_{2}^2,m_{3}^2\}$, with $m_{i}^2$ the eigenvalues of $M_\phi^2$. In the above three classes of solutions, we find:
\begin{eqnarray} \label{eq:scalar_mass}
\text{(1)}&&m_{1}^2=2 g_1 v^2\,,\quad
m_{2}^2=m_{3}^2= g_2 v^2\,;\nonumber\\
\text{(2)}&&m_{1}^2= 2 (3 g_1 + 2g_2) u^2\,,\quad
m_{2}^2=m_{3}^2=- 2 g_2 u^2\,;\label{eq:flavon_masses}\\
\text{(3)}&&m_{1}^2=2 (2 g_1 + g_2) s^2\,,\quad
m_{2}^2=-2m_{3}^2=-2 g_2 s^2\,,\nonumber
\end{eqnarray}
at $v_m$, $u_n$ and $s_l$, respectively. 
Although $m_{1}^2$ is always positive in all solutions, $m_{2}^2$ or $m_{3}^2$ may be positive or negative, depending on the sign of the coefficient $g_2$. 
The third solution class is less interesting since $m_{2}^2$ and $m_{3}^2$ always take opposite signs at $s_l$. 
Thus, $s_l$ is always an unstable saddle point of $V(\phi)$ and cannot be a vacuum. 
For the first two classes of solutions, if $g_2>0$, $m_{2}^2$ and $m_{3}^2$ are positive at $v_m$ and negative at $u_n$. 
Therefore, $u_n$ is an unstable saddle point, and $V(\phi)$ can only take a local minimum value (thus, the global minimum value) at $v_m$. 
As a result, $v_m$ is the only choice of the $\phi$ vacuum expectation value (VEV). 
On the contrary, if $g_2<0$, $u_n$ is the only choice of the $\phi$ VEV. 
All VEVs for $g_2 >0$ and $g_2 <0$ have been geometrically shown in Figs.~\ref{fig:DW_S} and \ref{fig:DW_T}, respectively, in the three-dimensional field space up to a mass-dimension normalisation factor. 


\section{Domain wall solutions \label{sec:DW}}
A domain wall (DW) is characterised by a classical and stable one-dimensional scalar field profile in three-dimensional space.
The equation satisfied by the scalar is obtained from the scalar Equation of Motion (EOM), i.e., the Euler-Lagrange equation of the scalar, which is given by
\begin{eqnarray}
 \Big(\frac{d^2}{dt^2} - \bigtriangledown^2 \Big) \phi_i + \frac{\partial V(\phi)} {\partial \phi_i} = 0 \,,
\end{eqnarray}
where $i$ denotes the components of the scalar $\phi$.
A DW refers to a stable solution of the scalar field and thus is time-independent. 
Defining the direction perpendicular to the DW surface as the $z$-axis, the profile variation is only $z$-dependent. 
Thus the equation of motion for the scalar field is simplified to
\begin{eqnarray}
\frac{d^2\phi_i(z)}{dz^2} = \frac{\partial V(\phi)} {\partial \phi_i} \,.
\label{eq:eom}
\end{eqnarray}
Given a theory where a discrete symmetry is spontaneously broken, a series of vacua (e.g., $v_1, v_2, ...$ as obtained above) can be generated. These vacua are disconnected from each other in the field space. By fixing two disconnected vacua $v_i$ and $v_j$ ($i\neq j$) {along the $z$-axis}, a DW can form between them. 
If the theory provides more than two vacua, there would be multiple choices {of} pairs of vacua, and different DWs can be generated.

In this section, we derive all DW solutions between any degenerate vacua discussed in Sec.~\ref{sec:vac}. 
Following Eq.~\eqref{eq:eom}, we refer to the spatial coordinate perpendicular to the wall's surface as $z$. 
Two different vacua are fixed at $z = \pm \infty$ as two different boundary conditions. 
The solution $\phi_i(z)$ varies continuously from one vacuum, (e.g., $v_1$ at $z= + \infty$) to another vacuum (e.g., $v_2$ at $z = -\infty$). 
At any $z$ coordinate between two boundaries, $\phi_i(z)$ does not take a vacuum value, and thus, there must be some energy stored in the scalar field. 
The total energy density stored in all scalars compared with the vacuum is a sum of the gradient energy and the potential
\begin{eqnarray} \label{eq:energy_density}
\rho(z) = \sum_i \frac{1}{2} \big[\phi_i'(z) \big]^2 + \Delta V(\phi(z)) \,,
\end{eqnarray}
where the potential is shifted such that the vacuum energy is zero, i.e., $\Delta V(\phi(z)) = V(\phi(z)) - V(\phi(\overline{z}\to \infty))$. 
Integrating along $z$ gives
the tension of the wall, i.e., the energy per unit area of the wall, 
\begin{eqnarray}
\sigma = \int_{-\infty}^{+\infty} dz \, \rho(z)\,.
\end{eqnarray}
The DW has a thickness $\delta$ representing the length scale over which most of the energy is localised. 
Following the discussion in \cite{Wu:2022stu}, we define the thickness $\delta$ as\footnote{In the special $Z_2$ DW case, the width $\delta$ is defined as the factor appearing in the hyperbolic tangent function of the profile $\propto \tanh(z/\delta)$ \cite{Vilenkin:1984ib}. This leads to the condition
\begin{eqnarray}
\int_{-\delta/2}^{\delta/2} dz \, \rho(z) = \frac{1}{2} \tanh \frac12 \Big(3-\tanh ^2 \frac12\Big) \times \sigma \approx 64.38\% \times \sigma \,,
\end{eqnarray} 
where $\sigma$ is the wall tension. We apply this condition as a generalised definition of the wall thickness for walls despite a hyperbolic profile.}
\begin{eqnarray}
\delta = \int_{z_0-\delta/2}^{z_0+\delta/2} dz \, \rho(z) \approx 64.38\% \times \sigma \,.
\end{eqnarray}
This definition applies to the case where $\rho(z)$ has only a single peak {centred} at $z=z_0$, which refers to the core of the DW. Without loss of generality, we set $z_0 = 0$ when solving scalar profiles perpendicular to the DW. 

\subsection{Domain walls separating $Z_2$-preserving vacua: S-type}\label{sec:DWS}
We first discuss DWs formed in separating $Z_2$-preserving vacua $v_m$. {It is straightforward to prove that each vacua is invariant under the transformation of $S$ or its conjugate transformations. Due to this, we denote these kinds of DWs in general as S-type DWs.}
It is convenient to normalise the fields and the $z$-coordinate as 
\begin{eqnarray}
\overline{\phi}_i = \frac{\sqrt{g_1}}{\mu} \phi_i = \frac{\phi_i}{v}\,,\quad
\overline{z} = \mu \, z\,,
\end{eqnarray} 
and define $\beta = g_2/g_1$.
Then, the EOM simplifies to
\begin{eqnarray} \label{eq:EOM}
\frac{d^2\overline{\phi}_i(\overline{z})}{d\overline{z}^2} = \overline{\phi}_i [-1 + \overline{\phi}_1^2 + \overline{\phi}_2^2 + \overline{\phi}_3^2 + \beta (\overline{\phi}_j^2 + \overline{\phi}_k^2) ]\,,
\end{eqnarray}
which includes only one free parameter $\beta$. 
The DW tension is given by $\sigma = \mu v^2 \overline{\sigma}$ with dimensionless tension
\begin{eqnarray}
\overline{\sigma} = \int_{-\infty}^{+\infty} d\overline{z} \, \left\{ \frac{1}{2} \big[\overline{\phi}^{\prime 2}_1(\overline{z}) + \overline{\phi}^{\prime 2}_2(\overline{z}) + \overline{\phi}^{\prime 2}_3(\overline{z}) \big] + \Delta \overline{V}(\overline{\phi}(\overline{z})) \right\}\,,
\end{eqnarray}
where $\Delta \overline{V} = \Delta V/(\mu^2 v^2)$. 
The thickness of the DW can be parameterised into a dimensionless parameter as $\overline{\delta} = \mu \delta$. 
{The S-type DWs are further classified into two topologically different types of DWs, SI and SII.}
The topology of SI and SII DWs are shown in Fig.~\ref{fig:DW_S} and the profiles of the scalars in the field space for typical values of $\beta$ are shown in Fig.~\ref{fig:Path_S}.

\subsubsection*{SI domain walls \label{sec:SI}}
The first type of DWs includes those separating $v_1$ and $v_4$, $v_2$ and $v_5$, and $v_3$ and $v_6$.  We follow the convention in \cite{Wu:2022tpe} and denote them as 
\begin{equation}\label{eq:SIvac}
\mybox{v_1}\mybox{v_4}\,, \quad
\mybox{v_2}\mybox{v_5}\,, \quad
\mybox{v_3}\mybox{v_6}\,, 
\end{equation}
respectively. $\myybox{v_i}\myybox{v_j} = \myybox{v_j}\myybox{v_i}$ (for $i\neq j$) because exchanging $v_i$ and $v_j$ on the two boundaries gives a physically equivalent DW. 

\begin{figure}[t!]
\centering
\includegraphics[width=.32\textwidth]{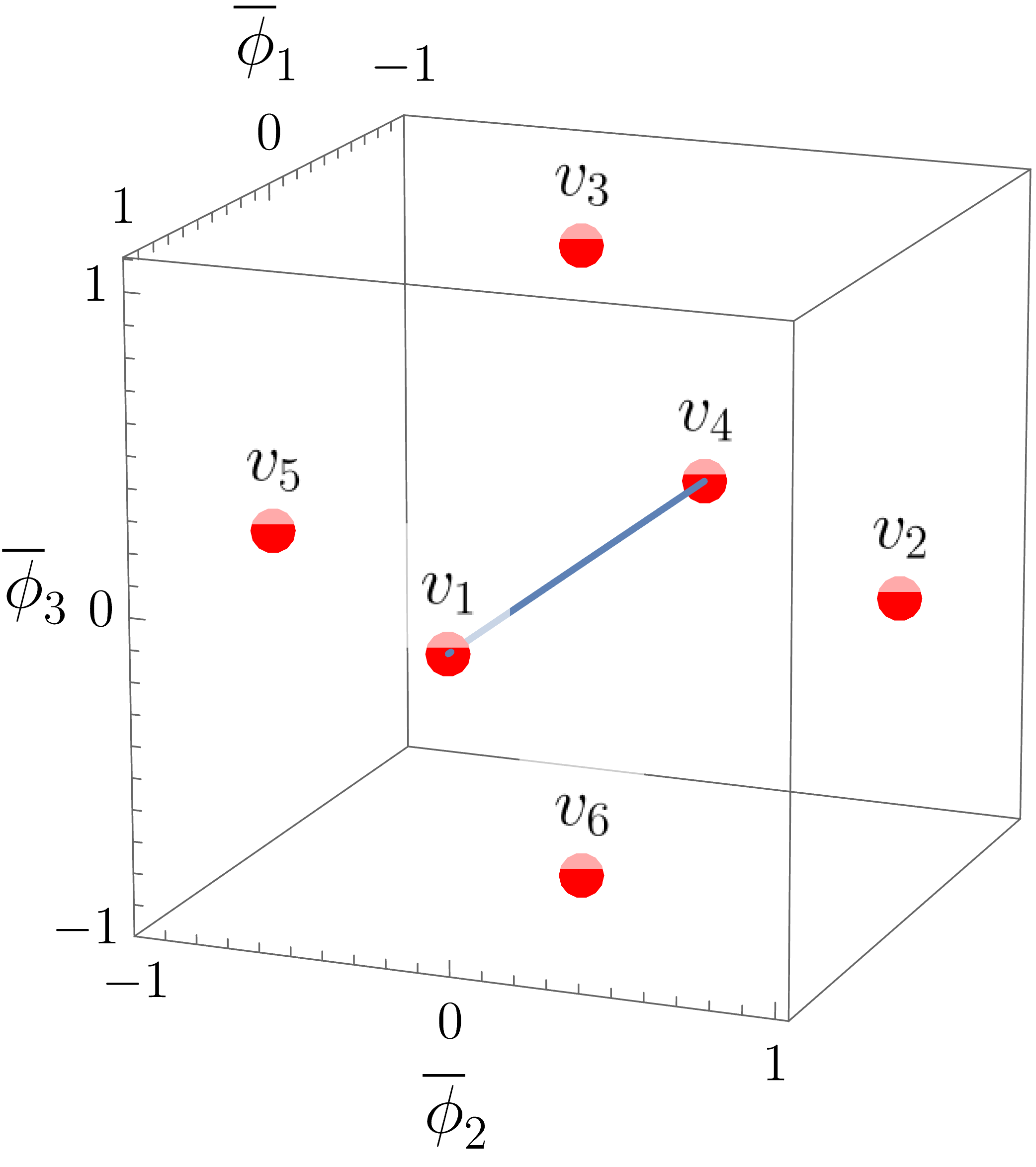} 
\hspace{2cm}
\includegraphics[width=.32\textwidth]{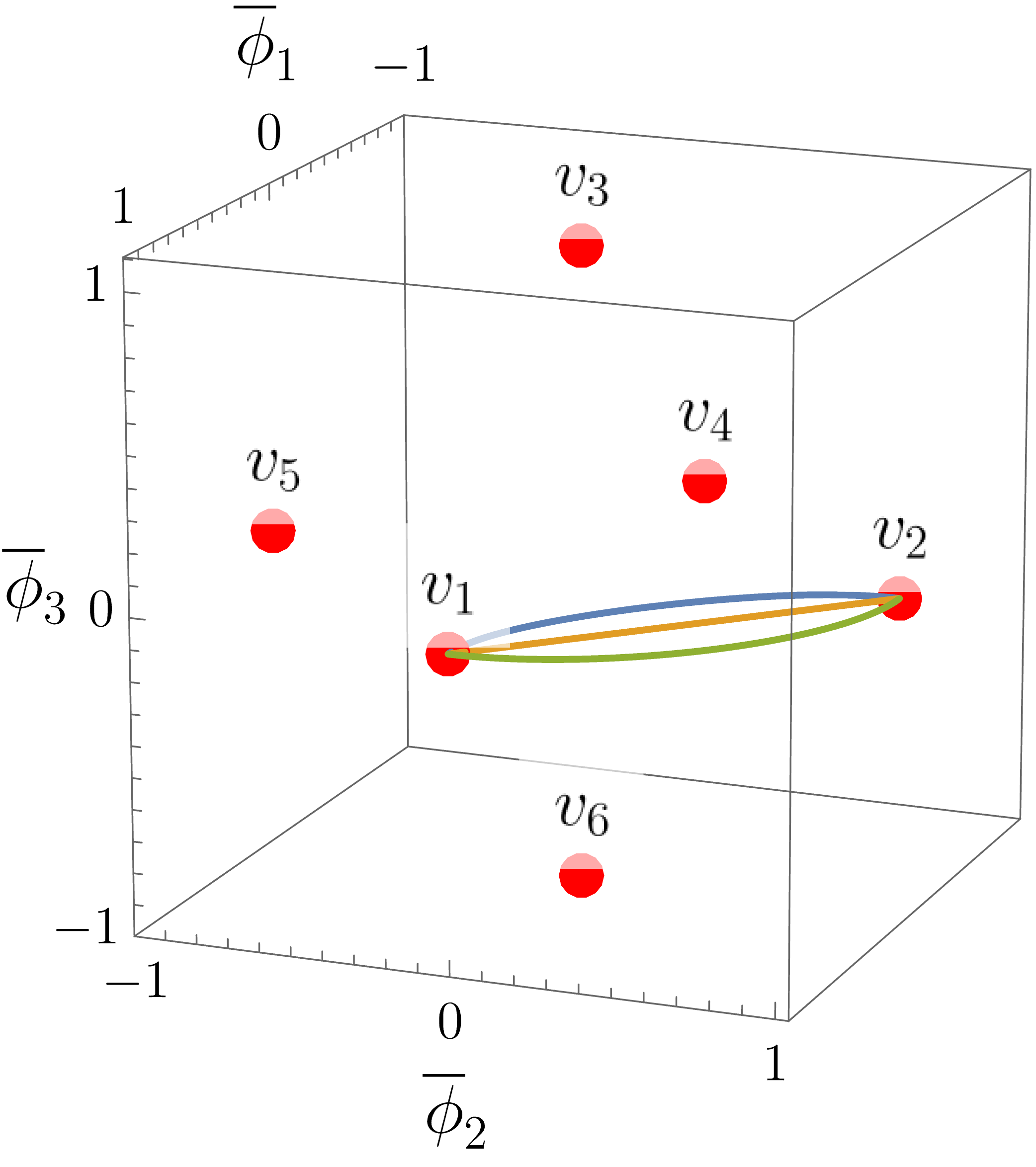}
\caption{Profiles of scalars in the field space for SI DW $\mybox{v_1}\mybox{v_4}$ (left panel) and SII DW $\mybox{v_1}\mybox{v_2}$ (right panel). The green, orange and blue curves refer to $\beta = 0.1,2,10$, respectively.\label{fig:Path_S}}
~\\
\centering
\includegraphics[width=.45\textwidth]{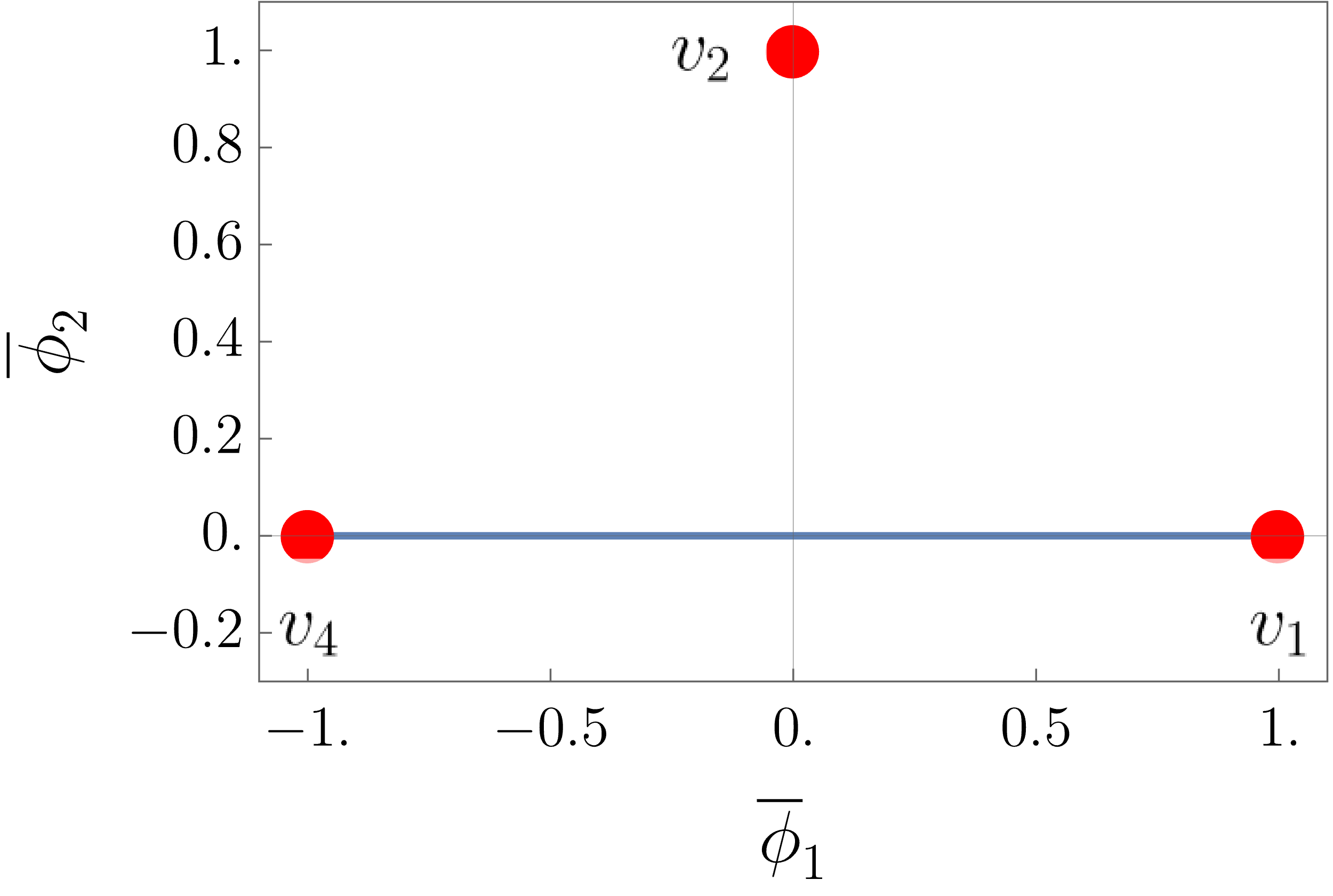}
\hspace{0.5cm}
\includegraphics[width=.45\textwidth]{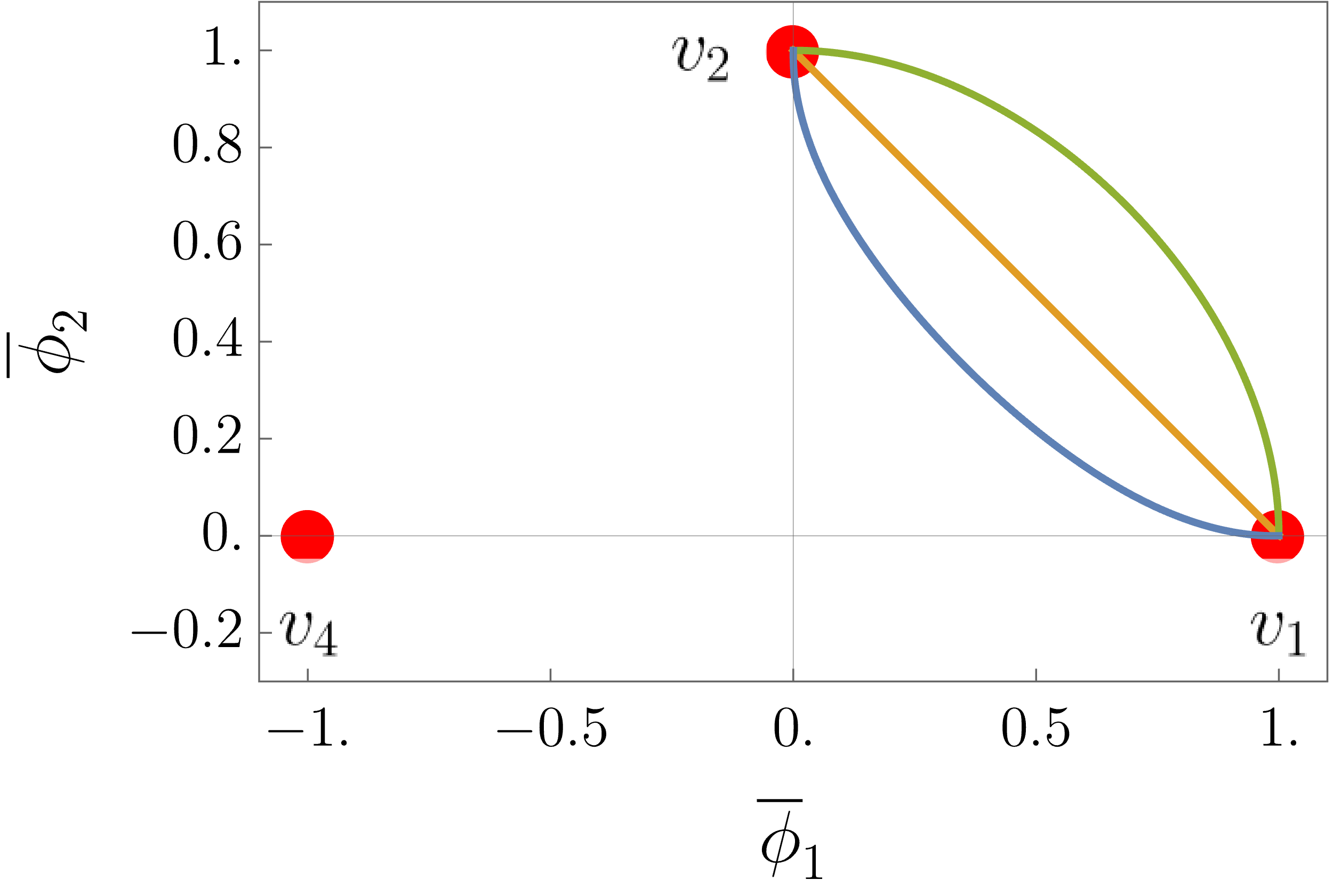} 
\caption{Projection of the paths in Fig.~\ref{fig:Path_S} in $\overline{\phi}_1-\overline{\phi}_2$ plane at $\overline{\phi}_3=0$.}
\label{fig:Path_S_slice}
~\\
\centering
\includegraphics[width=.49\textwidth]{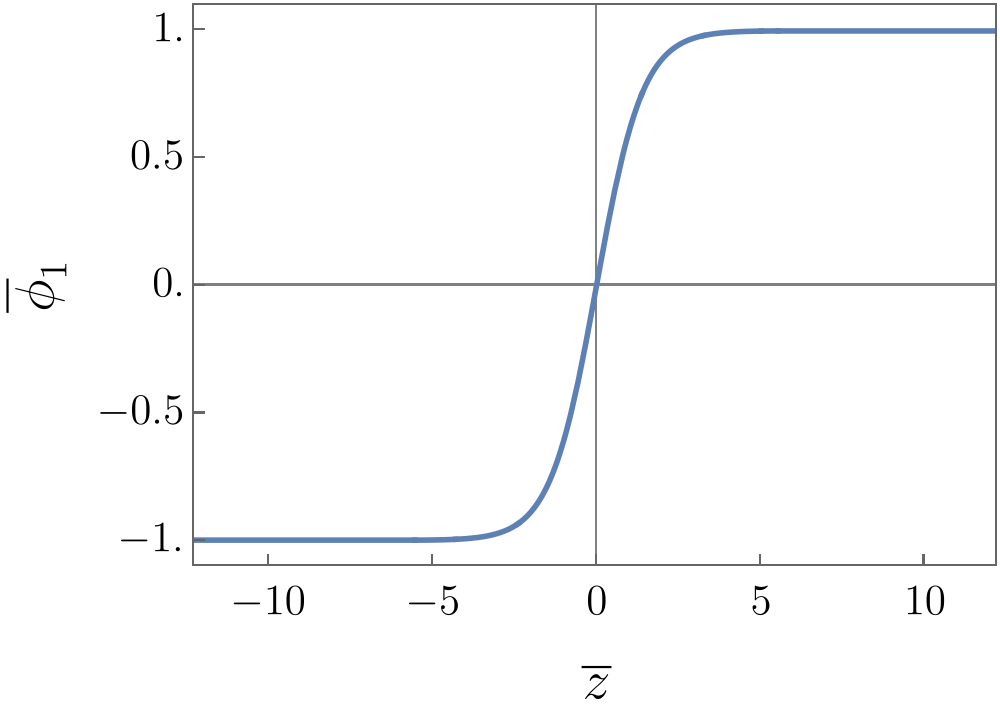} 
\includegraphics[width=.49\textwidth]{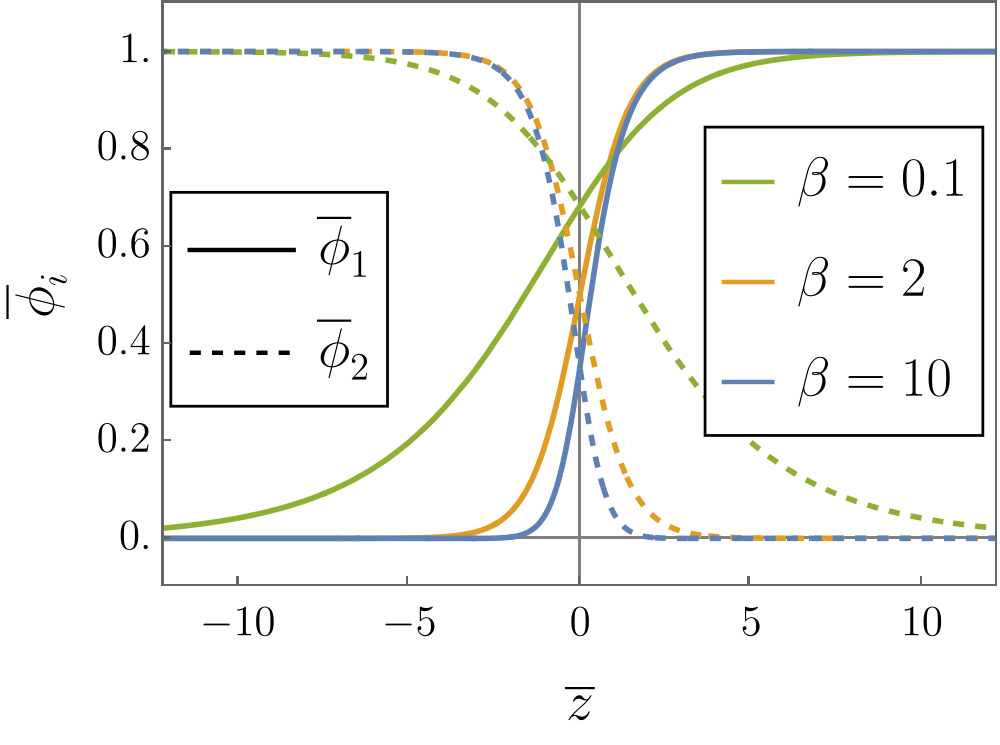} 
\caption{Profiles of scalars along the spatial coordinate $\overline{z}$ for SI DW $\mybox{v_1}\mybox{v_4}$ (left) and SII DW {$\mybox{v_1}\mybox{v_2}$} (right). $\overline{\phi}_2=\overline{\phi}_3=0$ along the path for SI DW $\mybox{v_1}\mybox{v_4}$, while $\overline{\phi}_3=0$ along the path for SII DW $\mybox{v_1}\mybox{v_2}$. The green, orange and red curves refer to $\beta = 0.1, 2, 10$, respectively.}
\label{fig:Profile_S}
\end{figure}
To make a quantitative analysis, we take $\mybox{v_1}\mybox{v_4}$ as an example. 
In this case, we find a DW solution that can be solved explicitly.
As shown in the left panel of Fig.~\ref{fig:Path_S} and more clearly as a projection in the left panel of Fig.~\ref{fig:Path_S_slice}, the solution is a straight line between $v_1$ and $v_4$ in the field space. 
When $\overline{\phi}_2$ and $\overline{\phi}_3$ are both fixed to be 0, the EOM of $\overline{\phi}_1$ and the boundary conditions are given by
\begin{eqnarray} \label{eq:EOM_SI}
&&\overline{\phi}_1''(\overline{z})= \overline{\phi}_1 [-1 + \overline{\phi}_1^2 ]\,, \nonumber\\
&&\overline{\phi}_1|_{\overline{z}\to +\infty} = +1\,, \quad \overline{\phi}_1|_{\overline{z}\to -\infty} = - 1 \,.
\end{eqnarray}
This is simply the DW formed from the spontaneous symmetry breaking of $Z_2$, which has been widely discussed in the literature. 
We refer to the latter case as the $Z_2$ DW.
The scalar field follows the profile 
\begin{eqnarray} \label{eq:profile_SI}
\overline{\phi}_1(\overline{z}) = \tanh \Big(\frac{\overline{z}}{\sqrt{2}}\Big)\,,
\end{eqnarray}
which is also shown in the left panel of Fig.~\ref{fig:Profile_S}. 
The denominator in the Hyperbolic function in Eq.~\eqref{eq:profile_SI} denotes the dimensionless thickness of the wall, i.e. $\overline{\delta} = \sqrt{2}$. 
The dimensionless tension can be straightforwardly calculated:
\begin{eqnarray}
\overline{\sigma}_{\text{SI}} = \int_{-\infty}^{+\infty} d\overline{z} \, \left\{ \frac{1}{2} \overline{\phi}^{\prime 2}_1(\overline{z}) + \frac{1}{4} \Big[\overline{\phi}^{2}_1(\overline{z})-1 \Big]^2 \right\} = \frac{2\sqrt{2}}{3} \,. \label{eq:tension_SI}
\end{eqnarray}
With the dimensionless thickness and tension, one can express the physical tension and thickness in terms of physical parameters as 
\begin{eqnarray}
\sigma_{\text{SI}} = \frac{2}{3} \sqrt{2g_1} v^3 = \frac23 m_1 v^2 \,, \quad \delta_{\text{SI}} = \sqrt{\frac{2}{g_1 v^2}} = \frac{2}{m_1} \,,
\end{eqnarray}
where $m_1$ is the scalar mass given in Eq.~\eqref{eq:scalar_mass}. 
The result is independent of $\beta$ and consistent with the classical one for $Z_2$ DW \cite{Vilenkin:1984ib}. 

\subsubsection*{SII domain walls} 
The vacua combinations of the second type DWs are all listed below,
\begin{equation}\label{eq:SIIvac}
\begin{aligned}
&\mybox{v_1}\mybox{v_2}\,,~
 \mybox{v_1}\mybox{v_3}\,,~
 \mybox{v_1}\mybox{v_5}\,,~
 \mybox{v_1}\mybox{v_6}\,,~
 \mybox{v_4}\mybox{v_2}\,,~
 \mybox{v_4}\mybox{v_3}\,,\\[6pt]
&\mybox{v_4}\mybox{v_5}\,,~
 \mybox{v_4}\mybox{v_6}\,,~
 \mybox{v_2}\mybox{v_5}\,,~
 \mybox{v_2}\mybox{v_6}\,,~
 \mybox{v_3}\mybox{v_5}\,,~
 \mybox{v_3}\mybox{v_6}\,,
\end{aligned}
\end{equation}
which are presented visually in the right panel of Fig.~\eqref{fig:DW_S}.
We take $\mybox{v_1}\mybox{v_2}$ as an example to solve the scalar profiles perpendicular to the DW. At both $v_1$ and $v_2$, $\phi_3=0$ is satisfied, and the reflection symmetry $\phi_3 \leftrightarrow - \phi_3$ is satisfied on the boundaries. We remind the reader that the EOM in Eq.~\eqref{eq:EOM} also satisfies this reflection symmetry. Thus, we expect there is the DW solution obtained by fixing $\phi_3 \to 0$. Then, the EOM and boundary conditions are simplified to be
\begin{eqnarray} \label{eq:EOM_SII}
&&\overline{\phi}_1''(\overline{z}) = \overline{\phi}_1 [-1 + \overline{\phi}_1^2 + \overline{\phi}_2^2 + \beta \overline{\phi}_2^2 ]\,, \nonumber\\
&&\overline{\phi}_2''(\overline{z}) = \overline{\phi}_2 [-1 + \overline{\phi}_1^2 + \overline{\phi}_2^2 + \beta \overline{\phi}_1^2 ]\,, \nonumber\\
&&\left.\begin{pmatrix}\overline{\phi}_1 \\ \overline{\phi}_2 \end{pmatrix}\right|_{\overline{z}\to +\infty} = \begin{pmatrix} 1 \\ 0 \end{pmatrix}, \quad 
\left.\begin{pmatrix}\overline{\phi}_1 \\ \overline{\phi}_2 \end{pmatrix}\right|_{\overline{z}\to -\infty} = \begin{pmatrix} 0 \\ 1 \end{pmatrix} \,.
\end{eqnarray}
The tension simplifies to
\begin{eqnarray} \label{eq:Tension_SII}
\overline{\sigma}_{\text{SII}} = \int_{-\infty}^{+\infty} d\overline{z} \, \left\{ \frac{1}{2} \Big[\overline{\phi}^{\prime 2}_1(\overline{z}) + \overline{\phi}^{\prime 2}_2(\overline{z})\Big] + \frac{1}{4} \Big[\overline{\phi}^{2}_1(\overline{z}) + \overline{\phi}^{2}_2(\overline{z})-1 \Big]^2 + \frac{\beta}{2} \overline{\phi}^{2}_1(\overline{z}) \overline{\phi}^{2}_2(\overline{z}) \right\} \,.
\end{eqnarray}
With the above parametrisation, scalar profiles and the dimensionless DW tension depend only on a single parameter, $\beta \equiv g_2/g_1$. 
Given any positive value of $\beta$, we can calculate the scalar profiles in the domain wall solution and the corresponding dimensionless tension and thickness numerically. 
The path of the scalars in the field space is shown in the right panel of Fig.~\ref{fig:Path_S} and as a projection in the right panel of Fig.~\ref{fig:Path_S_slice}. 
The domain wall solution is a straight line between $v_1$ and $v_2$ in the field space when $\beta=2$. {We note that an enhanced symmetry for $\beta = 2$ occurs since the potential of \equaref{eq:potential} becomes only a function of a single coupling. We expand this point further in \appref{app:B}.}
For $\beta<2$, $\phi_2$ is a concave function of $\phi_1$ and the path curves away from the origin; 
for $\beta>2$, in contrast, $\phi_2$ is a convex function of $\phi_1$ and the path moves toward the origin.
The profiles of $\overline{\phi}_2$ along $\overline{z}$ for $\beta = 0.1, 2, 10$ are shown in the right panel of Fig.~\ref{fig:Profile_S}, illustrating an increase in the dimensionless thickness as $\beta$ decreases.

\begin{figure}[t!]
\centering
\includegraphics[width=.48\textwidth]{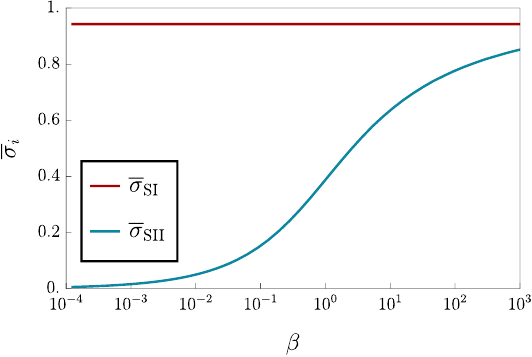} 
\includegraphics[width=.48\textwidth]{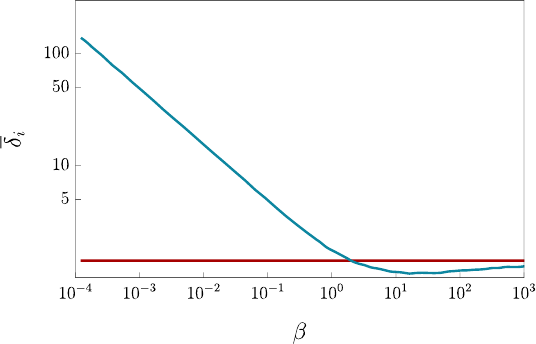} 
\caption{Dimensionless tension (left) and thickness (right) of DWs SI and SII as functions of $\beta$. For DW SI, $\overline{\sigma}_{\text{SI}} = 2\sqrt{2}/3$ and $\overline{\delta}_{\text{SI}} = \sqrt{2}$ are independent of $\beta$.}
\label{fig:Tension_S}
\end{figure}

To study the properties of the domain walls in detail, we perform a scan on $\beta$ from $10^{-3}$ to $10^3$ and obtain the dependence of dimensionless tension and thickness upon $\beta$, which is shown as the blue curves in Fig.~\ref{fig:Tension_S}. 
While the dimensionless tension of SI is independent of $\beta$ as computed in Eq.~\eqref{eq:tension_SI}, that of SII increases as $\beta$ becomes larger, as shown in the left panel of Fig.~\ref{fig:Tension_S}. 
Similarly, the dimensionless thickness of SI remains a constant for all $\beta$, but that of SII shows a negative correlation first and then a positive one with $\beta$.
The result can be semi-analytically expressed as
\begin{eqnarray} \label{tension_SII_ana}
\overline{\sigma}_{\text{SII}} (\beta) &\approx& 
\frac{2 \sqrt{2}}{3} \frac{1}{1+ 1.875 \beta^{-1/2}} \Big[ 1 + 0.5 \frac{\beta^{1/2}}{1+2 \beta} \Big] \,, \nonumber\\
\overline{\delta}_{\text{SII}} (\beta) &\approx& \frac{1.52}{\beta ^{0.50}}+\frac{0.66 \beta ^{0.79}}{\beta ^{0.70}+1}\,.
\end{eqnarray}
with relative error less than 6\% for $10^{-4} < \beta <10^3$. 

\subsection{Domain walls separating $Z_3$-preserving vacua: T-type}

\begin{figure}[t!]
\centering
\includegraphics[width=.32\textwidth]{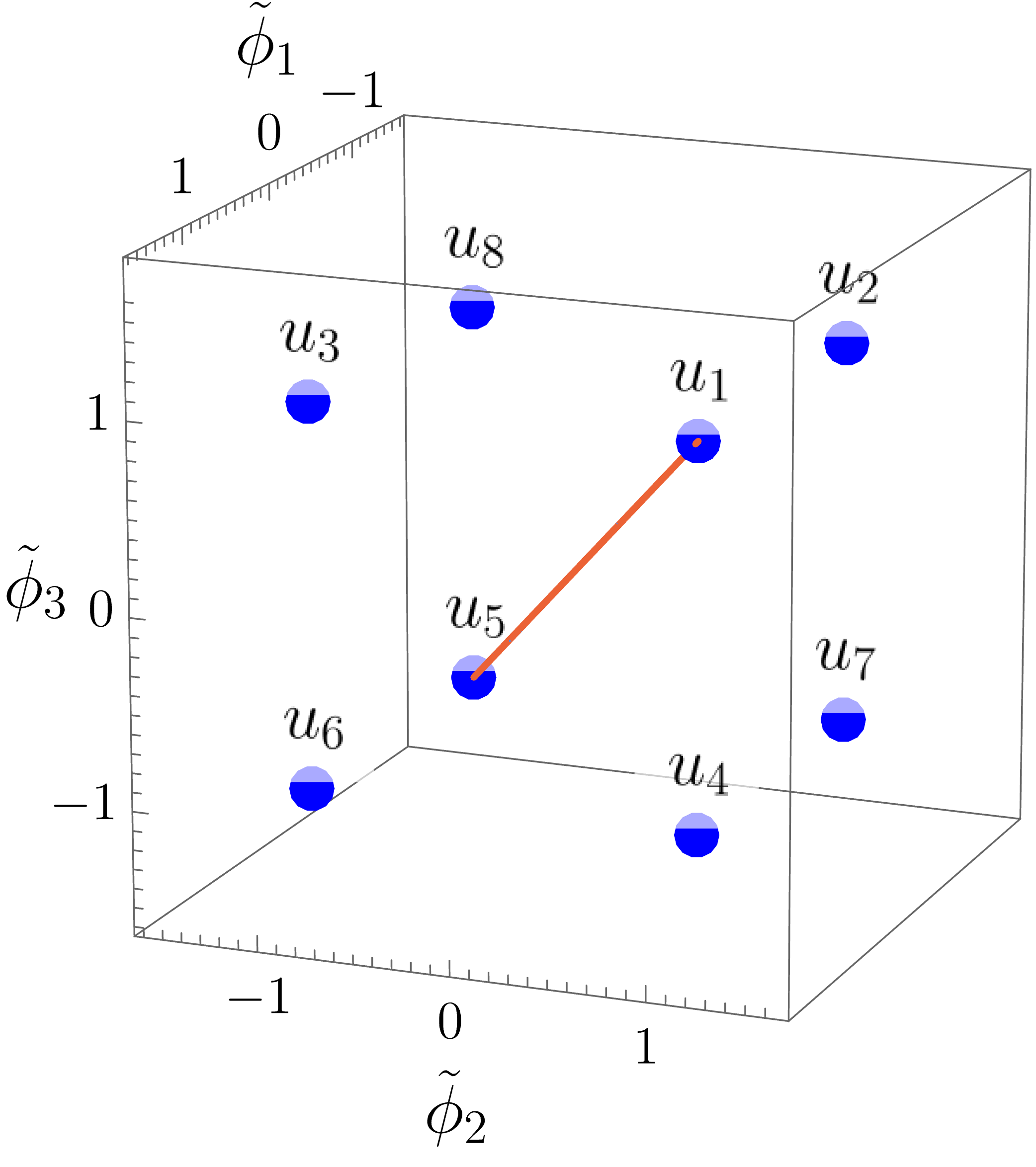}
\includegraphics[width=.32\textwidth]{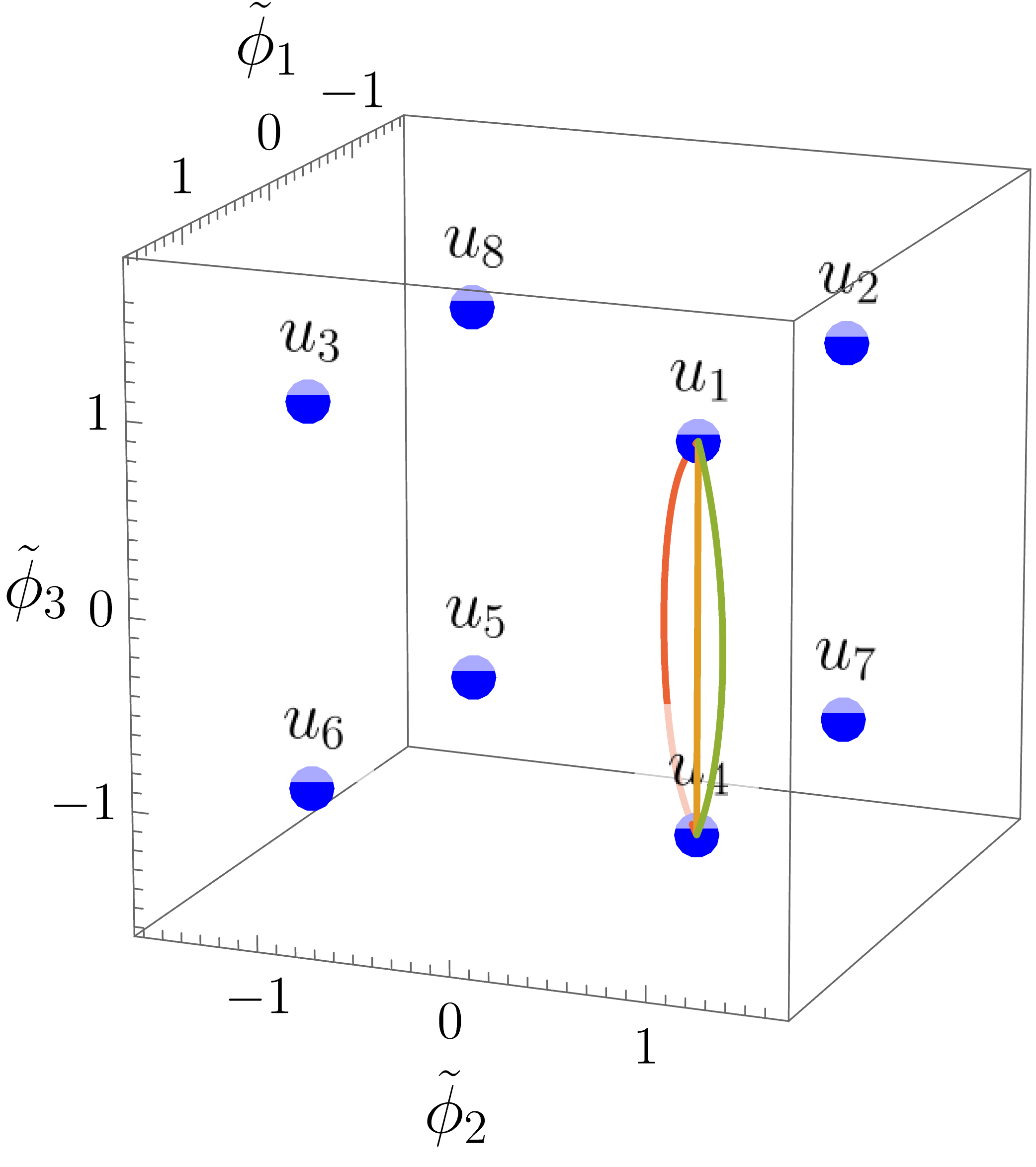}
\includegraphics[width=.32\textwidth]{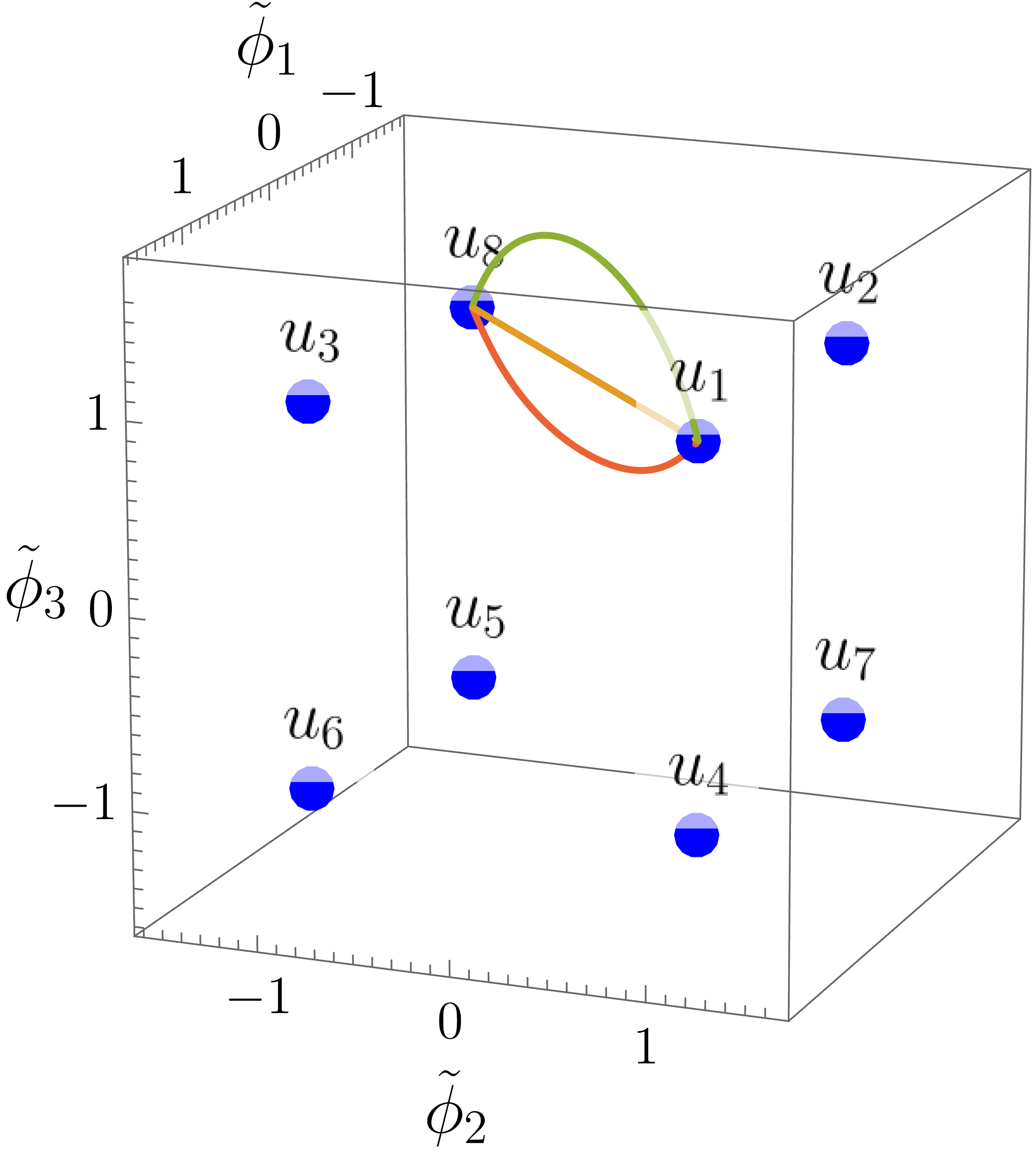}
\caption{Profiles of scalars in the field space for TI DW $\mybox{u_1}\mybox{u_5}$ (left panel), TII DW $\mybox{u_1}\mybox{u_4}$ (middle panel) and TIII DW $\mybox{u_1}\mybox{u_8}$ (right panel). The green, orange and red curves refer to $-\beta =0.1, 1, 1.4$, respectively.}
\label{fig:Path_T}
~\\
\centering
\includegraphics[width=.32\textwidth]{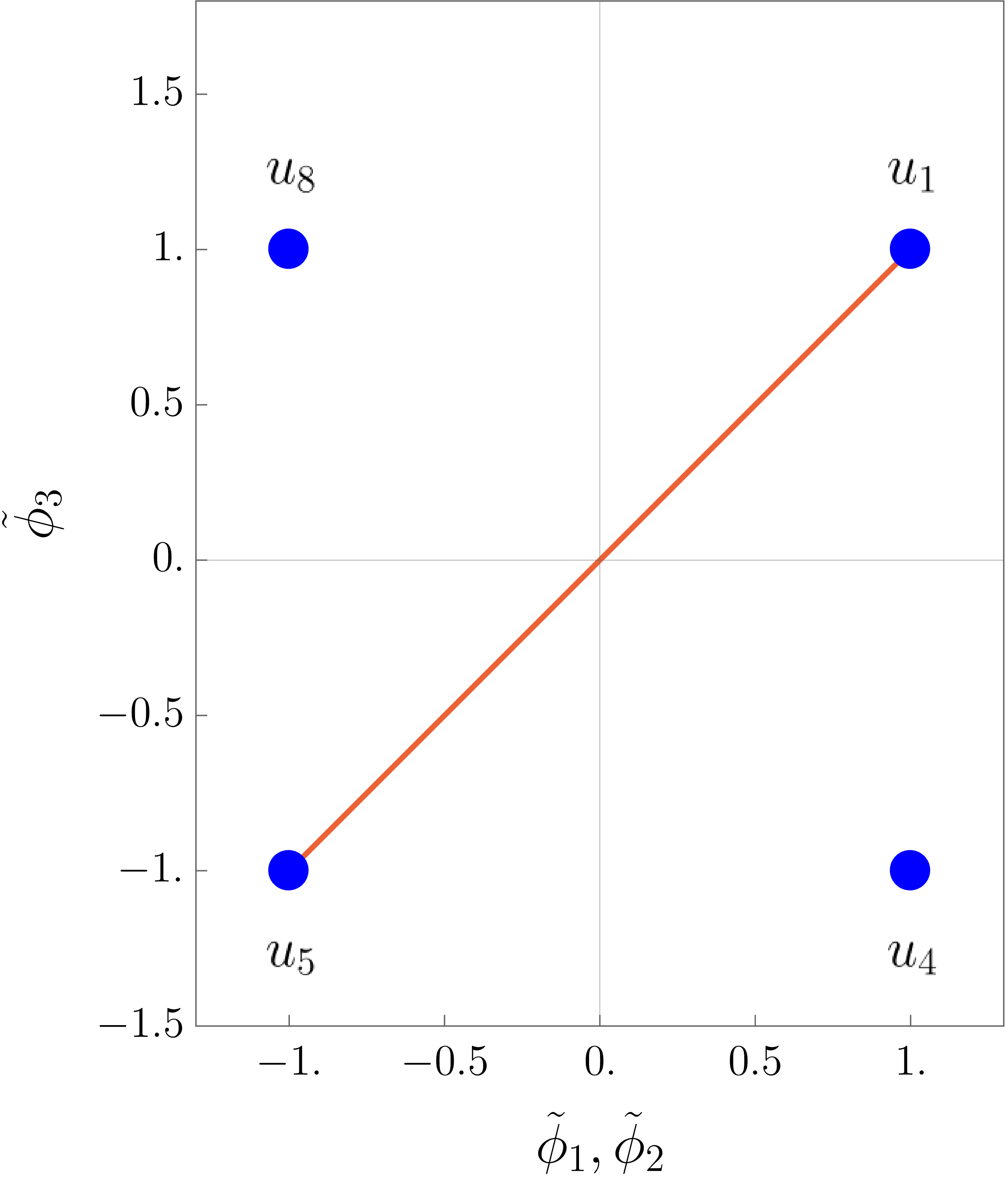}
\includegraphics[width=.32\textwidth]{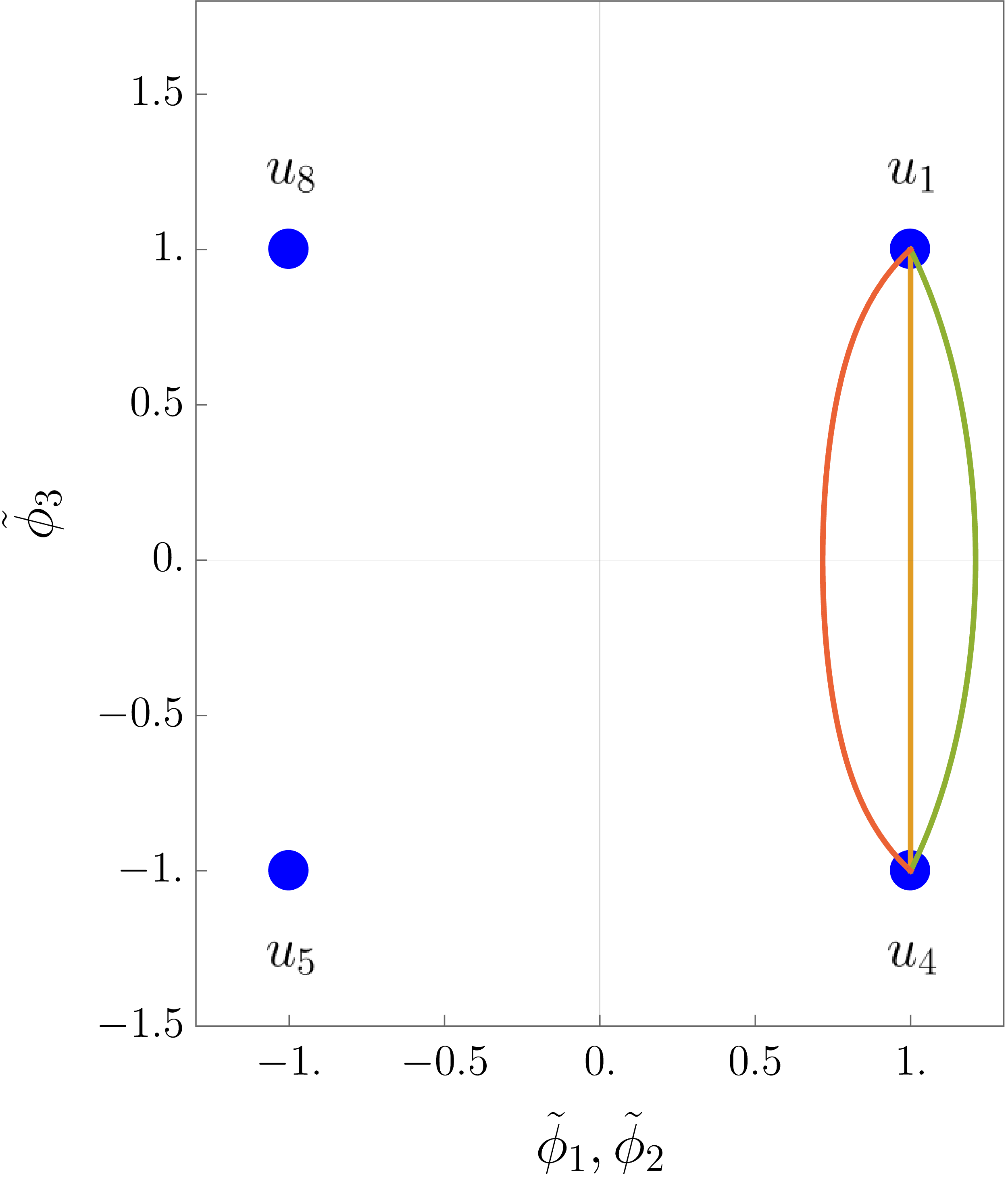}
\includegraphics[width=.32\textwidth]{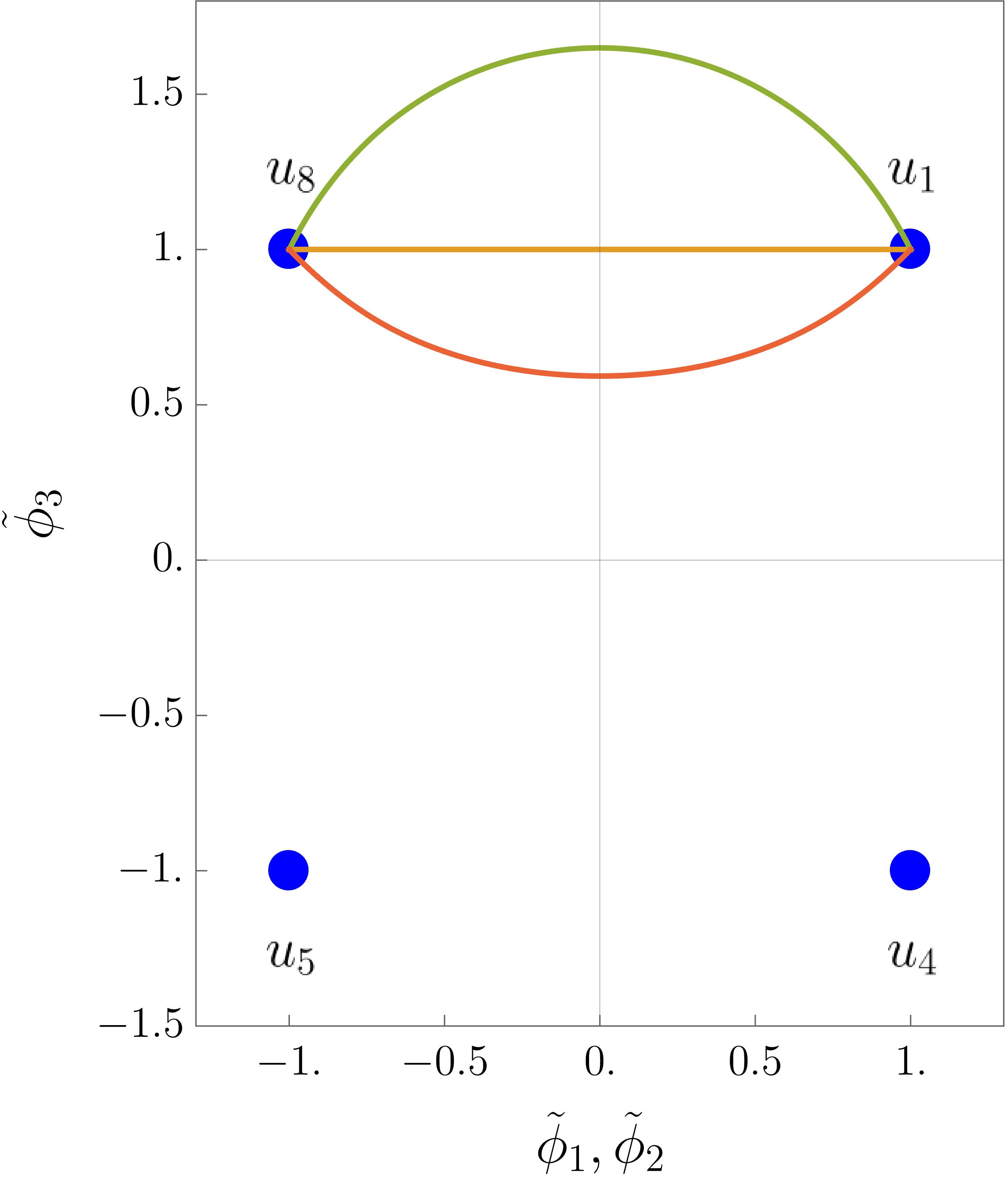} 
\caption{Projection of the paths in Fig.~\ref{fig:Path_T} in $\tilde{\phi}_1-\tilde{\phi}_3$ with $
\tilde{\phi}_1 = \tilde{\phi}_2$. }
\label{fig:Path_T_slice}
~\\
\centering
\includegraphics[width=.325\textwidth]{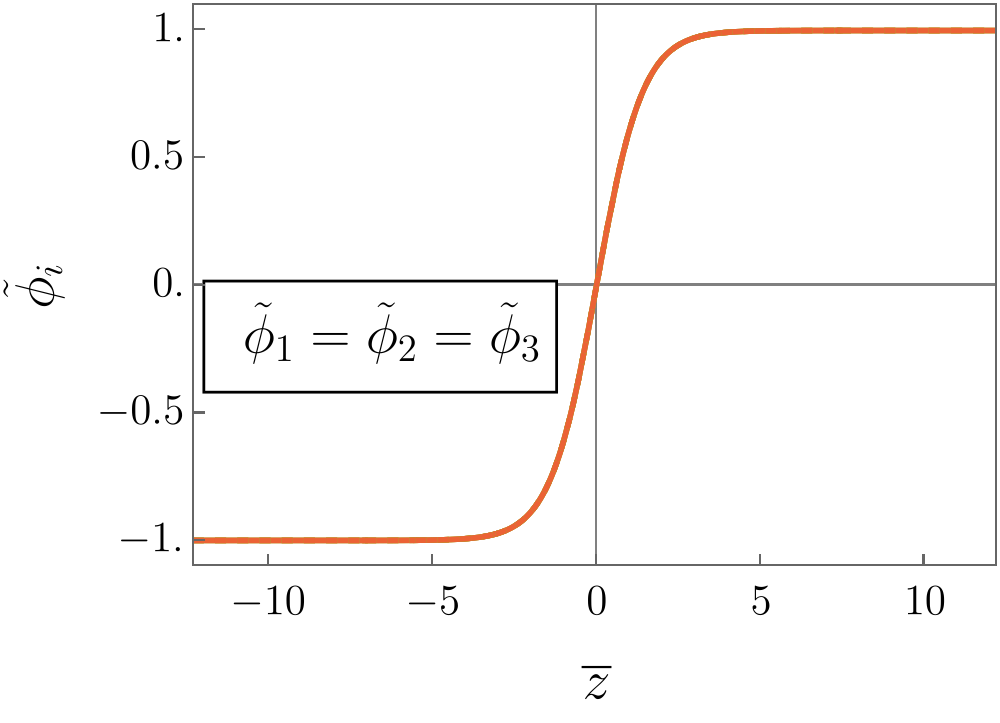} 
\includegraphics[width=.325\textwidth]{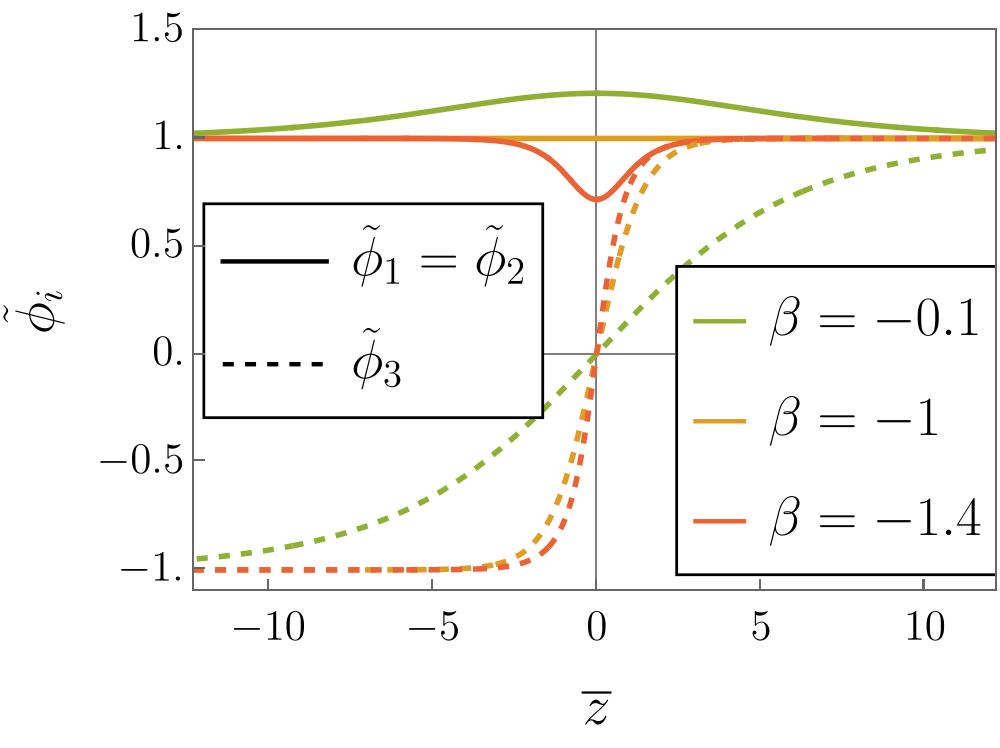} 
\includegraphics[width=.325\textwidth]{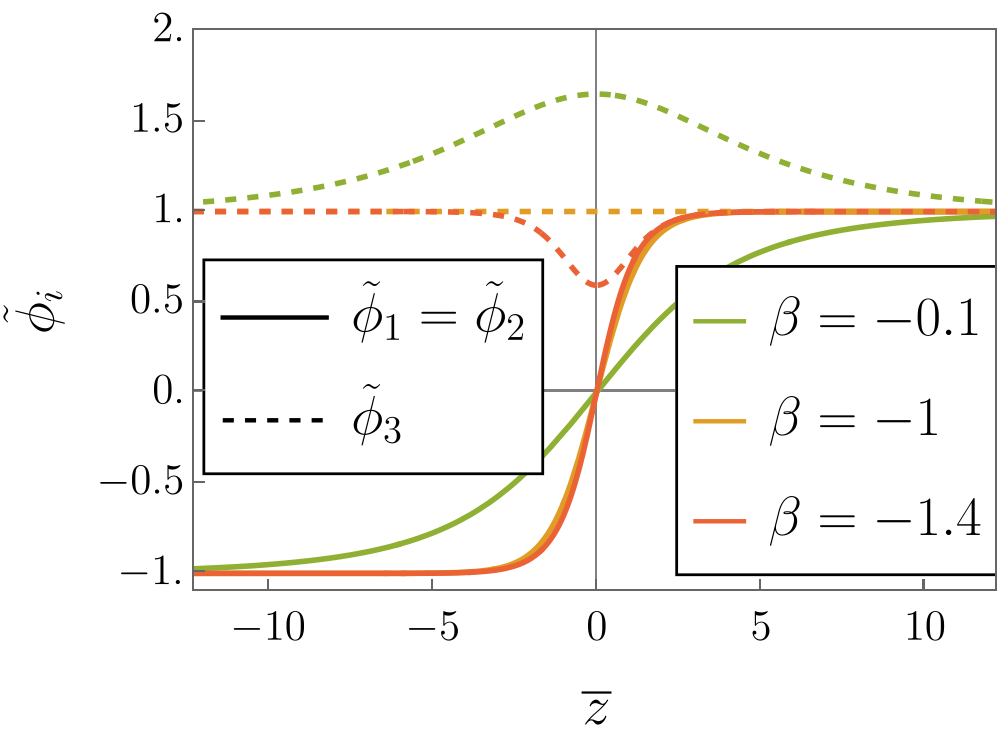} 
\caption{Profiles of scalars along the spatial coordinate $\overline{z}$ for TI DW $\mybox{u_1}\mybox{u_5}$ (left panel), TII DW $\mybox{u_1}\mybox{u_2}$ (middle panel) and TIII DW $\mybox{u_1}\mybox{u_8}$ (right panel). There are $\tilde{\phi}_1=\tilde{\phi}_2=\tilde{\phi}_3$, $\tilde{\phi}_2=\tilde{\phi}_3$, and $\tilde{\phi}_1=\tilde{\phi}_2$ along the path for the three cases, respectively. The green, orange and red curves refer to $-\beta =0.1, 1, 1.4$, respectively.}
\label{fig:Profile_T}
\end{figure}

The DWs separating $Z_3$-preserving vacua have three topologically different types, which are labelled as TI, TII and TIII, respectively. 
The label ``T'' refers to the vacua preserving residual $Z_3$ symmetries either generated by $T$ or its conjugate elements. 
For convenience, we perform the following normalisation for the fields and the coordinate
\begin{eqnarray}
\tilde{\phi}_i = \frac{\sqrt{3g_1+2g_2}}{\mu} \phi_i = \frac{\phi_i}{u}\,,\quad
\overline{z} = \mu \, z.
\end{eqnarray} 
Under such a normalisation, the EOM can be simplified into
\begin{eqnarray}
\frac{d^2\tilde{\phi}_i(\overline{z})}{d\overline{z}^2} = \tilde{\phi}_i \Big[-1 + \frac{1}{3+2\beta} ( \tilde{\phi}_1^2 + \tilde{\phi}_2^2 + \tilde{\phi}_3^2) + \frac{\beta}{3+2\beta} (\tilde{\phi}_j^2 + \tilde{\phi}_k^2) \Big]\,.
\end{eqnarray}
The DW tension is given by
$\sigma = \mu u^2 \tilde{\sigma} $ with
\begin{eqnarray}
\tilde{\sigma} = \int_{-\infty}^{+\infty} d\overline{z}\, \left\{ \frac{1}{2} \big[\tilde{\phi}_1^{\prime 2}(\overline{z}) + \tilde{\phi}_2^{\prime 2}(\overline{z}) + \tilde{\phi}_3^{\prime 2}(\overline{z}) \big] + \Delta \tilde{V}(\tilde{\phi}(\overline{z})) \right\} \,,
\end{eqnarray}
where $\Delta \tilde{V} = \Delta V/(\mu^2 u^2)$. The DW thickness is given by $\tilde{\delta} = \mu \delta$.
In particular, the EOMs become decoupled for each $\tilde{\phi}_i$ when $\beta=-1$, i.e. ${d^2\tilde{\phi}_i(\overline{z})}/{d\overline{z}^2} = \tilde{\phi}_i [-1 + \tilde{\phi}_i^2/(3+2\beta)]$. 
As a result, there is no correlation between the field when $\beta=-1$ and the paths in field space are straight lines.
The topology of TI, TII and TIII DWs are shown in Fig.~\ref{fig:DW_T}. 
Typical examples of scalar profiles in the field space are shown in Fig.~\ref{fig:Path_T} with $\beta = -0.1,\, -1,\, -1.4$, with projections in Fig.~\ref{fig:Path_T_slice}. 
Those profiles along the spatial coordinate are shown in Fig.~\ref{fig:Profile_T}. 
By varying the value of $\beta$, we obtain the $\beta$-dependent tensions and thicknesses of each type of DW, as shown in Fig.~\ref{fig:Tension_T}. 
We discuss each type of DWs in more detail in the following sections.

\subsubsection*{TI domain walls} 
Similar to the SI domain walls, the TI domain walls are also equivalent to the $Z_2$ domain walls, including
\begin{eqnarray}
\mybox{u_1}\mybox{u_5}\,,\quad
\mybox{u_2}\mybox{u_6}\,,\quad
\mybox{u_3}\mybox{u_7}\,,\quad
\mybox{u_4}\mybox{u_8}\,.
\end{eqnarray}
Again, we choose $\mybox{u_1}\mybox{u_5}$ as an example for detailed discussion. 
As visualised in the left panel of Fig.~\ref{fig:Path_T} and Fig.~\ref{fig:Path_T_slice}, the path of domain wall $\mybox{u_1}\mybox{u_5}$ in the field space is a straight line between $u_1$ and $u_5$. 
The EOMs and the boundary conditions for scalar fields across the DW are given by
\begin{eqnarray} \label{eq:EOM_TI}
&&\tilde{\phi}_1''(\overline{z}) = \tilde{\phi}_1 [-1 + \tilde{\phi}_1^2 ] \,, \nonumber\\
&&\tilde{\phi}_1 |_{\overline{z}\to + \infty} = 1\,,\quad
\tilde{\phi}_1 |_{\overline{z}\to - \infty} = - 1\,,
\end{eqnarray}
with $\tilde{\phi}_2 = \tilde{\phi}_3 = \tilde{\phi}_1$ satisfied. 
The EOM for $\tilde{\phi}_1$ and the boundary conditions are equivalent to Eq.~\eqref{eq:EOM_SI} and hence the profile of $\tilde{\phi}_1$ is in the same form as Eq.~\eqref{eq:profile_SI} and the dimensionless thickness of the wall is $\sqrt{2}$. 
The dimensionless tension includes contribution from all three fields $\tilde{\phi}_1,\tilde{\phi}_2, \tilde{\phi}_3$, with $2\sqrt{2}/3$ from each field equally. 
Thus the total energy stored in the DW is given by $3 \times 2\sqrt{2}/3 = 2\sqrt{2}$. 
We summarise the dimensionless tension and thickness as follows,
\begin{eqnarray}
\tilde{\sigma}_{\text{TI}} = 2 \sqrt{2} \,, \quad
\tilde{\delta}_{\text{TI}} = \sqrt{2} \,.
\end{eqnarray}

\subsubsection*{TII domain walls} 
The second type domain walls include
\begin{eqnarray}
&&\mybox{u_1}\mybox{u_2}\,,\quad
\mybox{u_1}\mybox{u_3}\,,\quad
\mybox{u_1}\mybox{u_4}\,,\quad
\mybox{u_2}\mybox{u_7}\,,\quad
\mybox{u_2}\mybox{u_8}\,,\quad
\mybox{u_3}\mybox{u_6}\,,\nonumber\\
&&\mybox{u_3}\mybox{u_8}\,,\quad
\mybox{u_4}\mybox{u_6}\,,\quad
\mybox{u_4}\mybox{u_7}\,,\quad
\mybox{u_5}\mybox{u_6}\,,\quad
\mybox{u_5}\mybox{u_7}\,,\quad
\mybox{u_5}\mybox{u_8}\,.
\end{eqnarray}
We take the DW $\mybox{u_1}\mybox{u_4}$ as an example. In both $u_1$ and $u_4$, $\tilde{\phi}_1$ and $\tilde{\phi}_2$ take the same value. 
Since there is no source to break the permutation symmetry between $\tilde{\phi}_1$ and $\tilde{\phi}_2$, we expect that $\tilde{\phi}_1 = \tilde{\phi}_2$ for $\overline{z}$ from $-\infty$ to $\infty$. In this case, EOM and boundary conditions for scalar profiles are simplified to 
\begin{eqnarray} \label{eq:EOM_TII}
&&\tilde{\phi}_1''(\overline{z}) = \tilde{\phi}_1 \Big[-1 + \frac{2+\beta}{3+2\beta}\tilde{\phi}_1^2+ \frac{1+\beta}{3+2\beta}\tilde{\phi}_3^2 \Big] \,, \nonumber\\
&&\tilde{\phi}_3''(\overline{z}) = \tilde{\phi}_3 \Big[-1 + \frac{2+2\beta}{3+2\beta}\tilde{\phi}_1^2+ \frac{1}{3+2\beta}\tilde{\phi}_3^2 \Big] \,, \nonumber\\
&&\left.\begin{pmatrix}\tilde{\phi}_1 \\ \tilde{\phi}_3 \end{pmatrix}\right|_{\overline{z}\to +\infty} = \begin{pmatrix} 1 \\ 1 \end{pmatrix}, \quad 
\left.\begin{pmatrix}\tilde{\phi}_1 \\ \tilde{\phi}_3 \end{pmatrix}\right|_{\overline{z}\to -\infty} = \begin{pmatrix} 1 \\ -1 \end{pmatrix} \,. 
\end{eqnarray}
By fixing $-\beta = 0.1, 1, 1.4$ respectively, we obtain solutions for scalar profiles in the field space in the middle panel of Fig.~\ref{fig:Path_T} and Fig.~\ref{fig:Path_T_slice}, and those along the $\overline{z}$ coordinate in the middle panel of Fig.~\ref{fig:Profile_T}. 
As mentioned, the path of the scalar profiles in the field space is a straight line between $u_1$ and $u_4$ when $\beta=-1$, which corresponds to the yellow line. 
The distance between the path and the origin positively depends on $\beta$. 

Following the numerical results of $\beta$-dependent dimensionless tension and thickness as the yellow lines in Fig.~\ref{fig:Tension_T}, we find that the dimensionless tension decreases as $\beta$ decreases while the dimensionless thickness increases. 
The following analytical formulas can fit the numerical results very well with relative error less than 3\%,
\begin{eqnarray}
\tilde{\sigma}_{\text{TII}}(\beta) = \frac{0.77 (-\beta)^{0.5}}{(1.5+\beta)^{0.25}} \,, \qquad
\tilde{\delta}_{\text{TII}}(\beta) = \frac{2.12 (-\beta)^{-0.5}}{0.48+(-\beta)^{1.35}} \,.
\end{eqnarray}

\subsubsection*{TIII domain walls} 

The third type of domain walls include
\begin{eqnarray}
&&\mybox{u_1}\mybox{u_8}\,,\quad
\mybox{u_1}\mybox{u_6}\,,\quad
\mybox{u_1}\mybox{u_7}\,,\quad
\mybox{u_6}\mybox{u_7}\,,\quad
\mybox{u_6}\mybox{u_8}\,,\quad
\mybox{u_7}\mybox{u_8}\,,\nonumber\\
&&\mybox{u_2}\mybox{u_3}\,,\quad
\mybox{u_2}\mybox{u_4}\,,\quad
\mybox{u_2}\mybox{u_5}\,,\quad
\mybox{u_3}\mybox{u_4}\,,\quad
\mybox{u_3}\mybox{u_5}\,,\quad
\mybox{u_4}\mybox{u_5}\,.
\end{eqnarray}
We take $\mybox{u_1}\mybox{u_8}$ as an example to solve for the DW solution. 
Similar to the TII case, there is no source to break the permutation symmetry between $\tilde{\phi}_1$ and $\tilde{\phi}_2$ and thus, we expect $\tilde{\phi}_1 = \tilde{\phi}_2$ for any $\overline{z}$. 
Then, the EOM and boundary conditions for $\tilde{\phi}_1$ and $\tilde{\phi}_3$ are given by:
\begin{eqnarray} \label{eq:EOM_TIII}
&&\tilde{\phi}_1''(\overline{z}) = \tilde{\phi}_1 \Big[-1 + \frac{2+\beta}{3+2\beta}\tilde{\phi}_1^2+ \frac{1 + \beta}{3+2\beta}\tilde{\phi}_3^2 \Big] \,,\nonumber\\
&&\tilde{\phi}_3''(\overline{z}) = \tilde{\phi}_3 \Big[-1 + \frac{2+2\beta}{3+2\beta}\tilde{\phi}_1^2+ \frac{1}{3+2\beta}\tilde{\phi}_3^2 \Big] \,,\nonumber\\
&&\left.\begin{pmatrix}\tilde{\phi}_1 \\ \tilde{\phi}_3 \end{pmatrix}\right|_{\overline{z}\to +\infty} = \begin{pmatrix} 1 \\ 1 \end{pmatrix}, \quad 
\left.\begin{pmatrix}\tilde{\phi}_1 \\ \tilde{\phi}_3 \end{pmatrix}\right|_{\overline{z}\to -\infty} = \begin{pmatrix} -1 \\ 1 \end{pmatrix} \,.
\end{eqnarray}
The scalar profiles in the field space obtained by numerical computation are shown in the right panel of Fig.~\ref{fig:Path_T} and Fig.~\ref{fig:Path_T_slice}.
Again, the path is straight between $u_1$ and $u_8$ for $\beta=-1$ and moves away from the origin as $\beta$ increases.
The scalar profiles along the $\overline{z}$ coordinate in the right panel of Fig.~\ref{fig:Profile_T}.
\begin{figure}[t!]
\centering
\includegraphics[width=.48\textwidth]{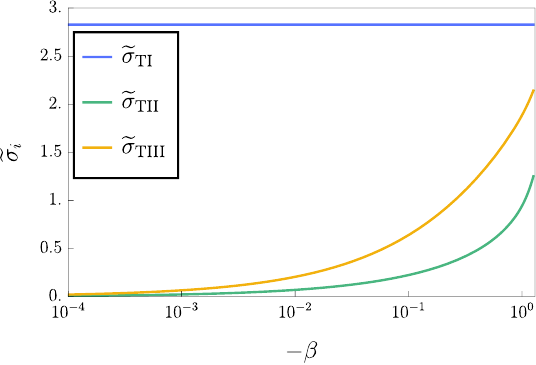} 
\includegraphics[width=.48\textwidth]{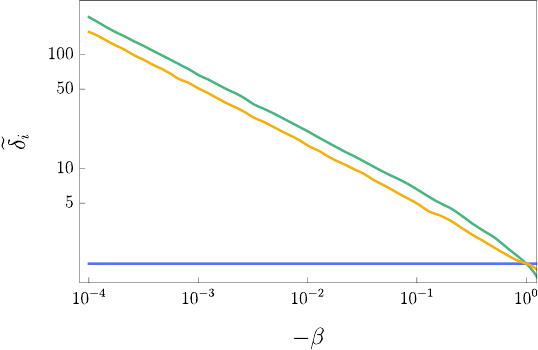} 
\caption{Dimensionless tension (left) and thickness (right) of DWs TI, TII, and TIII as functions of $\beta$. For DW TI, $\tilde{\sigma}_{\text{TI}} = 2\sqrt{2}$ and $\tilde{\delta}_{\text{TI}} = \sqrt{2}$ are independent of $\beta$. }
\label{fig:Tension_T}
\end{figure}
The dependence of dimensionless tension and thickness on the value of $\beta$ in the TIII case is qualitatively similar to the TII case, which is shown by the green lines in Fig.~\ref{fig:Tension_T}.
The following analytical formulas fit the numerical results with relative errors within 3\%,
\begin{eqnarray}
\tilde{\sigma}_{\text{TIII}}(\beta) &=& \frac{2.06 (-\beta)^{0.5}}{1+0.09 (-\beta)^{0.6}} \,, \nonumber\\
\tilde{\delta}_{\text{TIII}}(\beta) &=& \frac{1.62 (-\beta)^{-0.5}}{1+0.125 (-\beta)^{0.6}} \,.
\end{eqnarray}

\section{Stability of the domain wall solutions \label{sec:stability}}
Different DWs can be generated through the usual Kibble mechanism during phase transitions in the early universe. 
However, not all of them are necessarily stable. 
Given two vacua as boundary conditions and the generic static differential equations of scalars given in \equaref{eq:eom}, there may be other DW solutions beyond those in Sec.~\ref{sec:DW}.
These solutions, as shown below, could consist of several other DWs. 
If the energy stored in a DW solution, i.e., the tension, is not the lowest compared to other DW solutions, it can decay.
These unstable solutions with higher tensions can evolve into DWs with the lowest tension, i.e. the ground state. 
In this section, we determine the stability of S-type and T-type DWs separately.

\subsection{S-type domain walls}
\begin{figure}[t!]
\centering
\includegraphics[width=.4\textwidth]{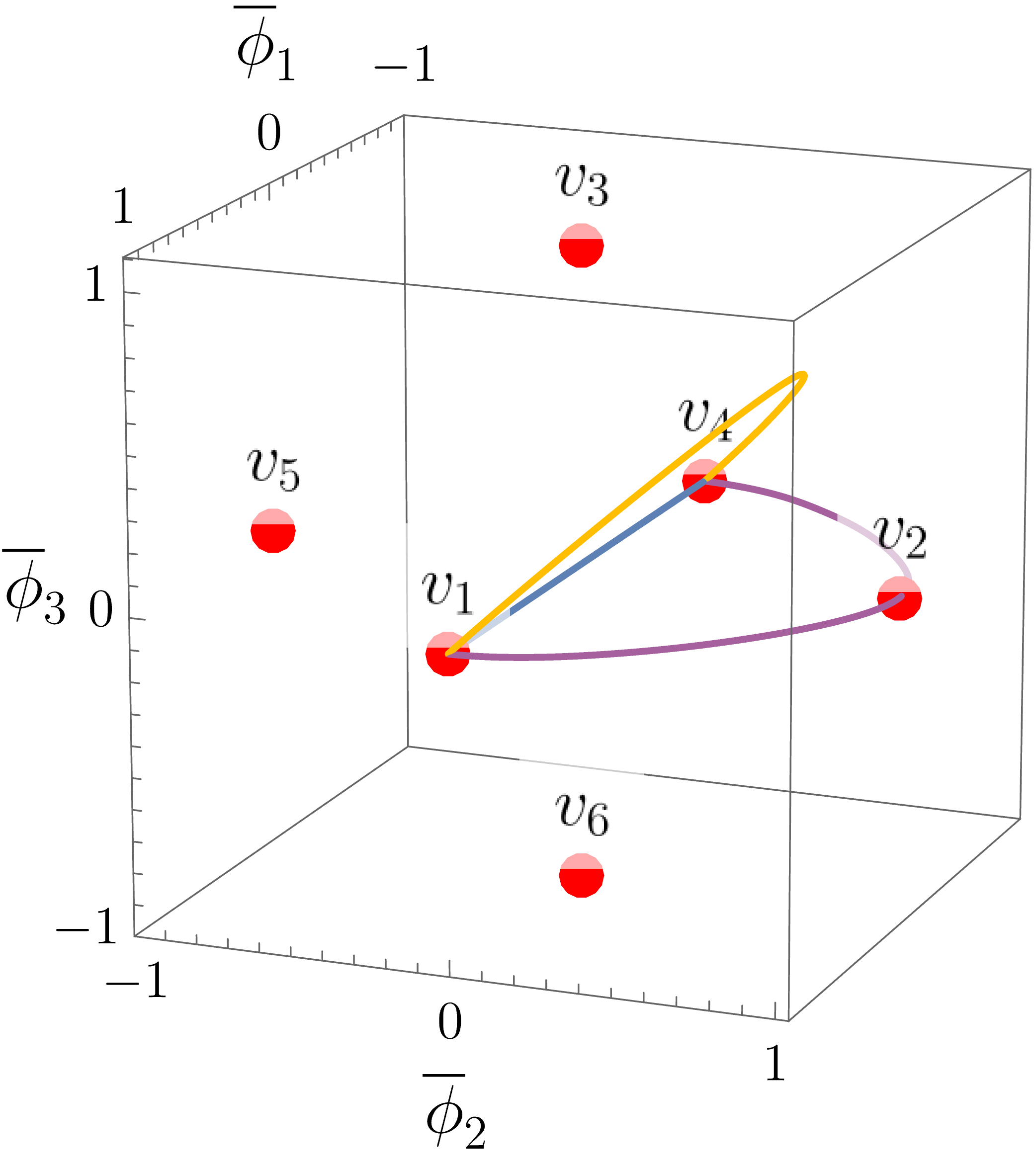} \hspace{1cm}
\includegraphics[width=.5\textwidth]{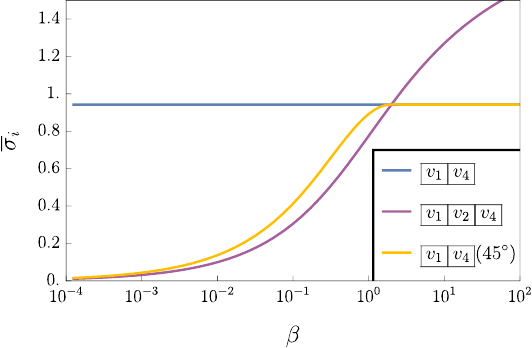} 
\caption{Left: Three different types of possible domain solutions from $v_1$ to $v_4$ when $\beta=0.3$. Right: The corresponding tension of the three different types of solutions for different $\beta$ values.}
\label{fig:v1v4}
\centering
\includegraphics[width=.45\textwidth]{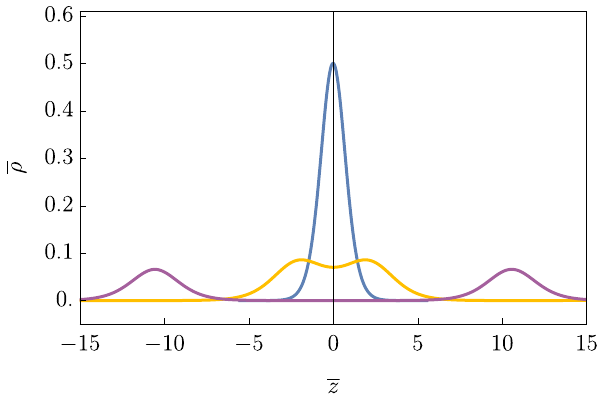} 
\caption{\label{fig:en_z} The energy density of the DW along the $z$ direction for the three different solutions in Fig.~\ref{fig:v1v4} with the same conventions.}
\end{figure}

In the last section, we summarised that there are two obviously distinct types of DWs, SI and SII, for the DWs separating two $ Z_2$-preserving vacua. 
To check the stability of the SI DW, we focus on the vacuum transition from $v_1$ to $v_4$. Given the EOM in Eq.~\eqref{eq:eom} and boundaries $v_1$ and $v_4$ at $z\to \pm \infty$, we will check if the SI DW solution, in which $v_1$ and $v_4$ are directly connected by a straight line in the field space, is the only solution. We have found several solutions, as shown in Fig.~\ref{fig:v1v4}:
\begin{enumerate}
\item The most obvious solution is the SI DW solution in Sec.~\ref{sec:SI}, e.g., $\mybox{v_1}\mybox{v_4}$, referring to the straight path connecting $v_1$ and $v_4$ in the field space. {We remind the reader that this solution is the $Z_2$ domain wall solution with the tension independent of the Lagrangian parameters $g_1$ and $g_2$.}
\item The second solution, {shown as the purple curve in Fig.~\ref{fig:v1v4}}, is a path indirectly between $v_1$ and $v_4$ with the third vacuum, e.g. $v_2$, as transit point. This path is a combination of two SII DWs $\mybox{v_1}\mybox{v_2}$ and $\mybox{v_2}\mybox{v_4}$ and we can denote the combination as $\mybox{v_1}\mybox{v_2}\mybox{v_4}$ for simplicity. Note that the transit point is an arbitrary choice for any vacuum of $v_2$, $v_3$, $v_5$ and $v_6$ and such paths in the field space constitute a class of paths which minimise the energy density of the domain wall for $\beta<2$.

\item A third solution is the straight line connection between $v_1$ and $v_4$ without crossing any other vacuum. However, the path in the field space is different from that of the SI solution but lies in the $45^{\circ}$ (or $135^{\circ}$, $225^{\circ}$, $315^{\circ}$) direction when projected to the $\overline{\phi}_2-\overline{\phi}_3$ plane, shown as orange curve in the left panel of Fig.~\ref{fig:v1v4}. 
\end{enumerate}
Tensions for different solutions as functions of the free parameter $\beta$ are presented in the right panel of Fig.~\ref{fig:v1v4}. 
The tension of the first solution, which is independent of $\beta$, is given by a constant $\overline{\sigma}_{\text{SI}} = \frac{2\sqrt{2}}{3}$. Tensions obtained in the other solutions are $\beta$-dependent. For $\beta<2$, tensions in these solutions are smaller than in the first solution, and the tension of the second solution is the smallest.
In this case, $\beta < 2$, $\bar{\sigma}_{\text{SI}} > 2 \bar{\sigma}_{\text{SII}}$ and thus the SI DW is unstable. For $\beta > 2$, the tension of the second solution is the largest, and the third solution gives tension equal to that of the first solution. In this case, $\beta > 2$,  $\bar{\sigma}_{\text{SI}}< 2 \bar{\sigma}_{\text{SII}}$ and thus the SI DW is stable.

When $\beta = 2$, the tensions of all three solutions are equal, which can be proved analytically. 
Notice that the second solution is equivalent to two SII DWs, i.e. $\mybox{v_1}\mybox{v_2}$ and $\mybox{v_2}\mybox{v_4}$. 
Proving the tension of the second solution to be $2\sqrt{2}/3$ is equivalent to proving that the tension of a SII DW at $\beta =2$ is $\sqrt{2}/3$, i.e., $\overline{\sigma}(\beta = 2) = \sqrt{2}/3$. 
With $\beta = 2$, 
it is convenient to make the parameterisation $\overline{\phi}_{\pm} = \overline{\phi}_1 \pm \overline{\phi}_2$ so that Eq.~\eqref{eq:EOM_SII} becomes 
\begin{eqnarray} \label{eq:EOM_SII_beta2}
&&\overline{\phi}_{\pm}''(\overline{z}) = \overline{\phi}_{\pm} [-1 + \overline{\phi}_{\pm}^2]\,, \nonumber\\
&&\left.\overline{\phi}_{\pm} \right|_{\overline{z}\to +\infty} = 1, \quad 
\left.\overline{\phi}_{\pm} \right|_{\overline{z}\to -\infty} = \pm 1 \,.
\end{eqnarray}
$\overline{\phi}_+$ in this case has a constant solution $\overline{\phi}_+(\overline{z})=1$ and $\overline{\phi}_-$ obeys the $Z_2$ DW solution, $\overline{\phi}_-(\overline{z})=\tanh(\frac{\overline{z}}{\sqrt{2}})$. Substituting this solution into Eq.~\eqref{eq:Tension_SII}, the tension is  
\begin{eqnarray}
\overline{\sigma}_{\text{SII}} (\beta = 2) = \frac{\sqrt{2}}{3} = \frac{1}{2} \overline{\sigma}_{\text{SI}}\,.
\end{eqnarray}
The SII DW does not have the stability problem as it always gives the smallest tension for vacuum transition between $v_1$ and $v_2$ for any positive value of $\beta$, which is obviously seen from the topology.

We show the dimensionless energy {density} distribution along the $z$ direction of the three different solutions between $v_1$ and $v_4$ in Fig.~\ref{fig:en_z} with $\beta = 0.3$, where 
\begin{eqnarray}
\overline{\rho}(\overline{z}) = \sum_i \frac{1}{2} \big[\overline{\phi}_i'(\overline{z}) \big]^2 + \Delta \overline{V}(\overline{\phi}(\overline{z}))\,,
\end{eqnarray}
is the dimensionless energy density normalised using Eq.~\eqref{eq:energy_density}. The blue curve refers to the energy density profile of SI DW $\mybox{v_1}\mybox{v_4}$. {As anticipated we observe that the energy density peaks at $\overline{z}=0$ as falls off at $|\overline{z}|\sim 1/0.3$.}
The purple curve shows the energy density profile of $\mybox{v_1}\mybox{v_2}\mybox{v_4}$ {and we see that there are two peaks (i.e. two domain walls) with the energy density peaking around $\overline{z}\sim \pm 10$ with magnitude significantly smaller than the SI DW scenario.} The orange curve refers to the third solution {where there is a flatter and broader domain wall than the SI DW case.}

The first and second solutions give tensions $\sigma_{\text{SI}} = 2\sqrt{2}/3$ and $2\sigma_{\text{SII}}(\beta = 0.3) \approx 0.5 $, respectively, representing the largest and smallest energy unit area stored in the scalar fields. In this case, the SI DW is unstable and {decays} to two SII DWs. We anticipate that during this decay the difference in energy of the higher and lower tension domain walls is released as gravitational radiation or possible light scalar (flavon) degrees of freedom.

We have numerically checked that given an initial SI path in the field space with a small perturbation to $\bar{\phi}_2$ and $\bar{\phi}_3$, the path automatically deviates from the initial one during the deformation iteration and eventually stabilised at the path $\mybox{v_1}\mybox{v_2}\mybox{v_4}$ or any other equivalent ones $\mybox{v_1}\mybox{v_{3,5,6}}\mybox{v_4}$. We developed code to numerically calculate domain wall solutions using a method similar to that of CosmoTransitions \cite{Wainwright:2011kj}, solving the equations of motion for scalar field configurations.
While CosmoTransitions focuses on solving the equations of motion for first-order phase transitions by finding bounce solutions and computing the Euclidean action, which determines the tunnelling probability between vacuum, our approach uses a similar methodology but for domain wall solutions.
{More specifically, we apply the path deformation algorithm implemented in CosmoTransition to solve Eq.~\eqref{eq:eom}, which is useful in solving a set of EOMs involving more than one field. For more details on the algorithm, see Ref.~\cite{Wainwright:2011kj}.
}

\subsection{T-type domain walls}
\begin{figure}[t!]
\centering
\includegraphics[width=.4\textwidth]{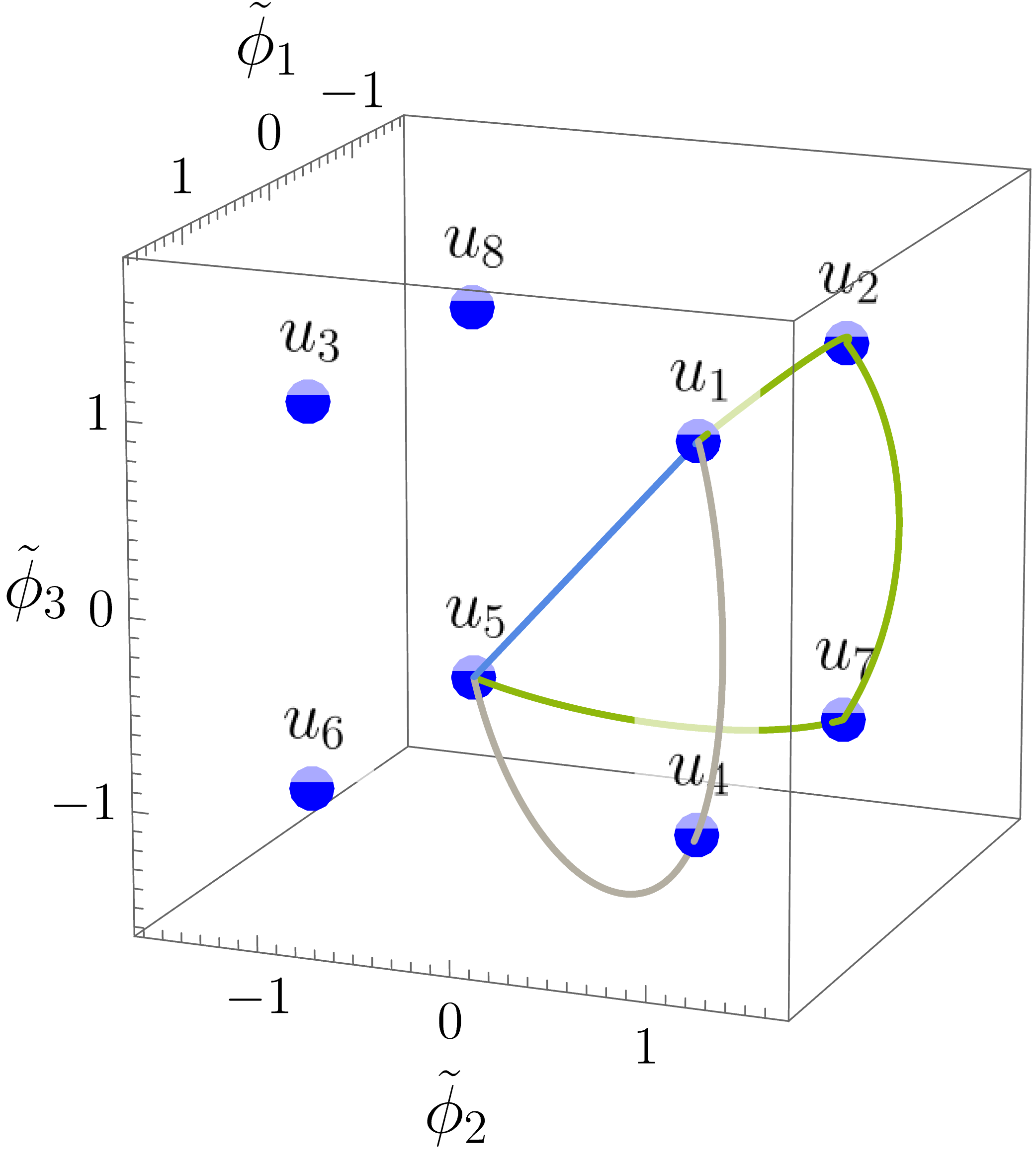} 
\includegraphics[width=.55\textwidth]{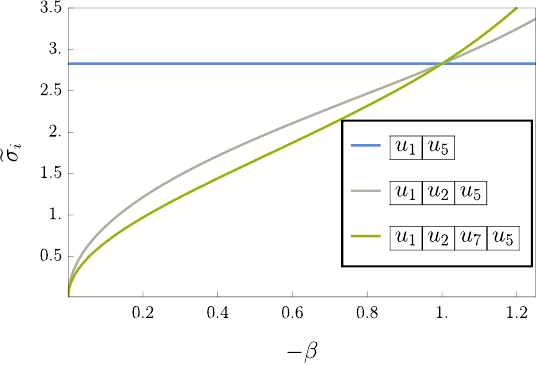} 
\caption{Left: Three different types of possible domain solutions from $u_1$ to $u_5$ when $\beta=-0.1$. Right: Tension of different possible DW solutions from $u_1$ to $u_5$.}
\label{fig:u1u5}
\end{figure}

For $\beta<0$, we arrive at three T-type DWs. The TII DW is always stable as seen from the topology.
We first check the stability of the TI DW. There can be three different types of solutions between $u_1$ and $u_5$, as shown in the left panel of Fig.~\ref{fig:u1u5} in field space. 
\begin{enumerate}
\item The TI DW which is a blue straight line in the left panel of Fig.~\ref{fig:u1u5} that runs between $u_1$ and $u_5$ in the field space and is labelled as $\mybox{u_1}\mybox{u_5}$. 
\item The combination of a TII DW and a TIII DW, which passes one of the other vacua in the field space, e.g. $\mybox{u_1}\mybox{u_4}$ and $\mybox{u_4}\mybox{u_5}$, or equivalently $\mybox{u_1}\mybox{u_4}\mybox{u_5}$. This is shown as a grey line in the left panel of Fig.~\ref{fig:u1u5}.
\item The combination of three TII DWs, which consists of three edges of the cube, e.g. $\mybox{u_1}\mybox{u_2}$, $\mybox{u_2}\mybox{u_7}$ and $\mybox{u_7}\mybox{u_5}$, or equivalently, $\mybox{u_1}\mybox{u_2}\mybox{u_7}\mybox{u_5}$. This is shown as a green line in the left panel of Fig.~\ref{fig:u1u5}.
\end{enumerate}
The tensions of the three different solutions are shown in the right panel of Fig.~\ref{fig:u1u5} for various $\beta$.
For $-\beta>1$, the TI DW, referred to as solution 1 here, has the lowest energy and is, therefore, energetically stable. The TI DW is unstable for $-\beta<1$ as solution 3 has the lowest energy. 
We then check the stability of the TIII DW. Taking $u_2$ and $u_5$ as boundary conditions. We find two types of solutions. 
\begin{enumerate}
\item 
One is the path directly connecting between $u_2$ and $u_5$, which is equivalent to that between $u_4$ and $u_5$. See $\mybox{u_4}\mybox{u_5}$ in Fig.~\ref{fig:u1u5}. 
\item The other is the path from $u_1$ to $u_2$ to $u_8$. This is equivalent to the path from $u_2$ to $u_7$ to $u_5$, i.e., $\mybox{u_2}\mybox{u_7}\mybox{u_5}$ in Fig.~\ref{fig:u1u5}. 
\end{enumerate}
By comparing tensions in these two solutions, we see that the TIII DW is stable when $-\beta>1$.
Finally, we conclude that both TI and TIII DWs are stable for $-\beta>1$ and unstable for $-\beta<1$.

\section{Gravitational waves from non-Abelian domain walls \label{sec:GW}}
In general, stable DWs can be problematic as their energy would dominate the total energy density of the universe. 
This problem can be solved if the residual discrete symmetry is softly broken by energy biases between the vacua \cite{Preskill:1991kd,Riva:2010jm,Ouahid:2018gpg}. This explicit symmetry breaking come from a number of mechanisms including from Planck suppressed operators or the chiral anomaly.
The energy biases can be expressed as
\begin{equation} 
V^{\mathrm{bias}}_{ij} = \epsilon_{ij} v^4\,,
\end{equation}
where $i,j$ refers to the scalar field indices with $v$ the VEV of the field in the absence of softly symmetry breaking. In addition to leading to the decay of domain walls, the bias term can help correct leading order mixing patterns which describe flavour mixing in the lepton sector \cite{Gelmini:2020bqg}.
These biases make the DWs unstable and annihilate at a temperature $T_{\mathrm{ann}}$, generating gravitational waves. 
The spectrum of the GW for frequency $f$ and at time $t$ is given by 
\begin{equation}
\Omega h^2(f,t) = \frac{h^2}{\rho_c(t)} \left(\frac{d \rho_{\mathrm{gw}}(t)}{d \ln f}\right)\,.
\end{equation}
For a typical $Z_2$ DW, the GW spectrum is characterised by a peak at the frequency 
\begin{align}
\begin{aligned} f_{\text {peak }} & \simeq a\left(t_{\mathrm{ann}}\right) H\left(t_{\mathrm{ann}}\right) \simeq 1 \times 10^{-4} \mathrm{~Hz}\left(\frac{g_*\left(T_{\mathrm{ann}}\right)}{10}\right)^{1 /6}\left(\frac{T_{\mathrm{ann}}}{\mathrm{TeV}}\right) \\ & \simeq 3 \times 10^3 \mathrm{~Hz}\left(\frac{10}{g_*\left(T_{\mathrm{ann}}\right)}\right)^{1 / 12}\left(\frac{V_{\mathrm{bias}}}{\sigma\, \mathrm{TeV}}\right)^{1 / 2}, 
\end{aligned}
\end{align}
with amplitude 
\begin{align}
\begin{aligned}
\left.\Omega_{\mathrm{gw}} h^2\right|_{\text {peak }} & \simeq 0.7 \times 10^{-37}\left(\frac{10}{g_*\left(T_{\text {ann }}\right)}\right)^{4 / 3}\left(\frac{\sigma}{\mathrm{TeV}\, T_{\text {ann }}^2}\right)^2 \\
& \simeq 0.9 \times 10^{-67}\left(\frac{10}{g_*\left(T_{\text {ann }}\right)}\right)^{1 / 3}\left(\frac{\sigma}{\mathrm{TeV}^3}\right)^4\left(\frac{\mathrm{TeV}^4}{V_{\text {bias }}}\right)^2,
\end{aligned}
\end{align}
where the present critical density $\rho_c(t_0) = 10.53 h^2 \mathrm{keV/cm^3}$ and effective degrees of freedom $g_{s*}(t_0) = 3.91$ is applied.
On each side of peak, we use the power-law approximation obtained from numerical simulation, which are $\Omega_{\mathrm{gw}} h^2 (f) \propto f^3$ for $f < f_\mathrm{peak}$ and $\Omega_{\mathrm{gw}} h^2 (f) \propto f^{-1}$ for $f > f_\mathrm{peak}$ 
\cite{Saikawa:2017hiv}.
In the non-Abelian DWs discussed in this work, some solutions are unstable even in the absence of any biases.
The evolution of those unstable DW solutions can change the kinetic history of the DW network and thus produce special gravitational waves signal. 
An interesting example is the decay of an SI DW into a pair of SII DWs, during which extra energy is released, partly into the kinetic energy of the walls and gravitational radiation. The details of those complex processes are beyond the scope of this work. Here, we focus on the case where all the wall solutions are stable, namely $\beta>2$ for the $Z_2$ preserving case and $\beta<-1$ for the $Z_3$ preserving case.

\subsection{Gravitational waves emission by non-Abelian domain walls }
For the S-type DWs, the SII DWs are four times more likely to be produced than SI DWs during the phase transition, see \equaref{eq:SIIvac} and \equaref{eq:SIvac}. 
We assume that the ratio of number density to be constant during the subsequent evolution and we have 
\begin{eqnarray*}
    n_{\rm SI}:n_{\rm SII} = 3:12\,,
\end{eqnarray*}
where 3 and 12 are numbers of vacuum combinations of SI and  SII DWs, respectively. 
To be consistent with cosmology, biases are introduced to make the DWs unstable and decay. 
From the vacuum structure in Fig.~\ref{fig:DW_S}, at first glance there are 15 biases between the six vacua: three for SI DWs and 12 for SII DWs. 
However, only five of them are linearly independent. 
Here, as a example, $v_1$ is chosen to be the lowest vacuum and the other five vacua are determined by the biases between them and $v_1$.
The biases for the benchmark case are summarised in Tab.~\ref{tab:epsilons}. 

\begin{table}[t!]
 \centering
 \begin{tabular}{|ccccc|}
 \hline\hline
 $\epsilon^v_{12}$ & $\epsilon^v_{13}$ & $\epsilon^v_{14}$ & $\epsilon^v_{15}$ & $\epsilon^v_{16}$ \\ \hline 
 $2 \hat \epsilon$ & $3 \hat \epsilon$ & $ \hat \epsilon$ & $4 \hat \epsilon$ & $5 \hat \epsilon$ \\\hline\hline
 \end{tabular}
 \begin{tabular}{|ccccccc|}
 \hline\hline
 $\epsilon^u_{12}$ & $\epsilon^u_{13}$ & $\epsilon^u_{14}$ & $\epsilon^u_{15}$ & $\epsilon^u_{16}$ & $\epsilon^u_{17}$ & $\epsilon^u_{18}$ \\
 \hline
 $2 \hat \epsilon$ & $4 \hat \epsilon$ & $6 \hat \epsilon$ & $\hat \epsilon$ &$ 3 \hat \epsilon$ & $5 \hat \epsilon$& $7 \hat \epsilon$ \\\hline\hline
 \end{tabular}
 \caption{Choice of the bias for the difference between the vacua of the $Z_2$ (left) and $Z_3$ (right) preserving scenario in terms of $\hat \epsilon$, which is a free parameter which we set when discussing testability. }
 \label{tab:epsilons}
\end{table}

\begin{figure}[t!]
\centering
\includegraphics[width=.48\textwidth]{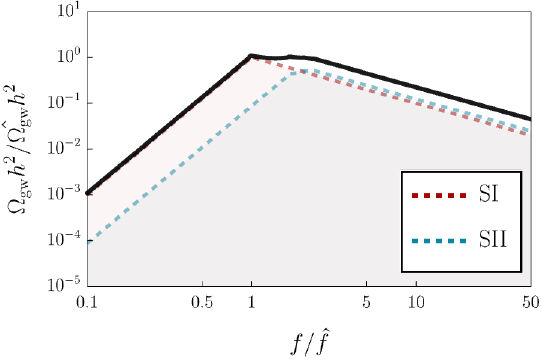} 
\includegraphics[width=.48\textwidth]{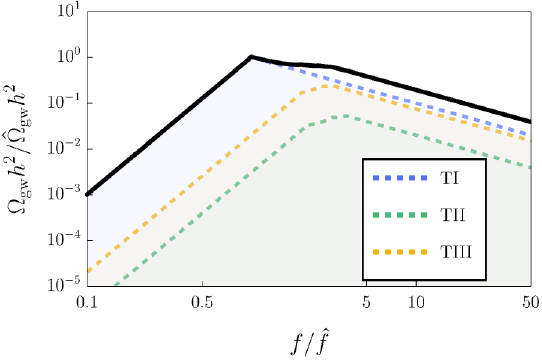} 
\caption{The colored dashed lines show the gravitational waves spectrum for each DW type. The black line shows the gravitational waves spectrum due to overlapping spectra for the S-type DWs (left) with $\beta = 10$, and T-type DWs (right) with $\beta = -1.2$.
\lm{The signal has been weighted according to the corresponding number density}, as discussed in the main text. The frequencies and the spectrum are normalised to $ \hat f = f^{Z_2}_{\mathrm{peak}}$, the peak frequency of the contribution from SI/TI DWs, and its correspondent amplitude $\hat \Omega h^2$. The values of five possible biases for the $Z_2$ case and seven possible biases for the $Z_3$ case are given in Tab.~\ref{tab:epsilons}. } 
\label{fig:GW_gnr}
\end{figure}

The left panel of Fig.~\ref{fig:GW_gnr} shows the resulting GW amplitude as a function of the wave-number. 
While the biases are free parameters, the tensions of the DWs can be fixed by the potential as discussed in Sec.\ref{sec:DW}.
The consequence of different possible tensions is similar to what has been observed in Ref.~\cite{Gelmini:2020bqg} but more significant. 
In Ref.~\cite{Gelmini:2020bqg}, the origin of having different peaks is the difference between the biases while the tension is unique. 
Here, since there are different types of DWs, these DWs with different tensions contribute to various peaks in the complete spectrum. 
This additional effect leads to a more pronounced multi-peak structure than the one caused by different biases. 
In particular, if $\Omega^{\rm SI} \sim \Omega^{\rm SII}$, the spectrum does not evolve as $f^{-1}$ in a significant interval of frequencies after the first peak, which is more likely to be detected.

For the T-type DWs, the ratio of the number densities of different vacua is 
\begin{eqnarray*}
    n_{\rm TI}:n_{\rm TII}:n_{\rm TIII}=4:12:12\,,
\end{eqnarray*}
where 4, 12 and 12 are numbers of vacuum combinations in S1, S2, S3, respectively.
Again, $u_1$ is chosen to be the lowest energy vacuum and the other seven vacua are determined by the biases between them and $u_1$.
The biases for the benchmark case are summarised in Tab.~\ref{tab:epsilons}. 
Following the same procedure as the S-type DWs, we find a similar behaviour, where there is a deviation from $f^{-1}$ power-law spectrum after the peak, as shown in the right panel of Fig.~\ref{fig:GW_gnr}. 

Finally, it is important to recall that the discussion above holds only when $\beta>2$ for the S-type DWs and $\beta<-1$ for the T-type DWs. If $\beta$ is smaller (larger) for the S-type (T-type) DWs, there would be unstable DWs which may decay much earlier than the annihilation time of the stable ones. 
Such a process may generate additional signals at higher frequencies and provide a different GW spectrum.

\subsection{Testability with gravitational wave observatories}
\begin{figure}[t!]
\centering
\includegraphics[width=.9\textwidth]{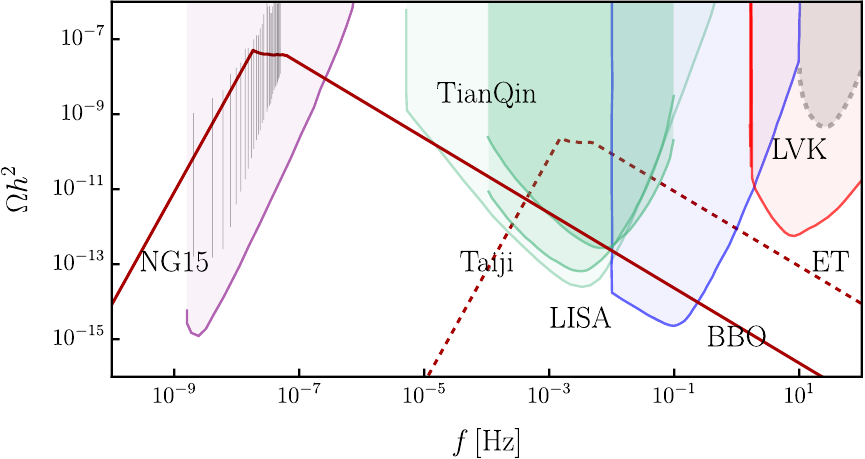} 
\includegraphics[width=.9\textwidth]{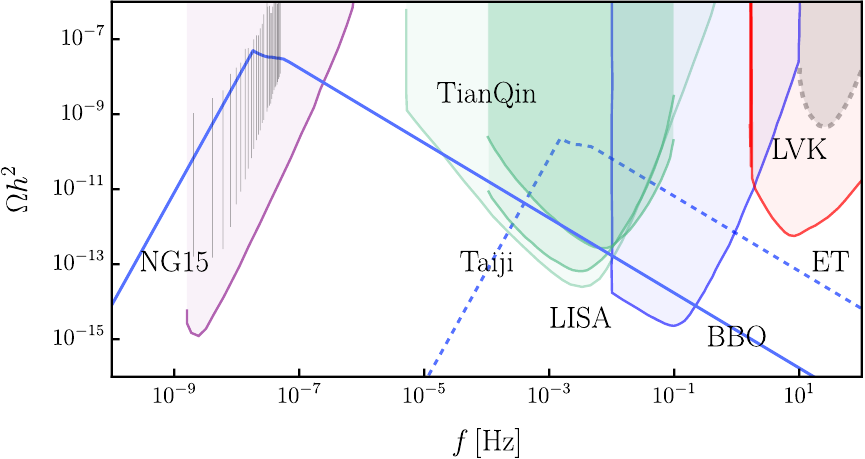} 
\caption{Gravitational waves spectrum due to overlapping spectra for the S-type DWs (top) with $\beta = 10$, and T-type DWs (bottom) with $\beta = -1.2$. The solid and dashed lines correspond to the benchmark points BP1 and BP2 in Tab.~\ref{tab:bps}, respectively.}
\label{fig:GW_sp}
\end{figure}
Given the recent discovery of a stochastic gravitational wave background by Pulsar Time Array (PTA) observatories and planned next-generation earth and space GW observatories, it is interesting to see how the deviation from the ordinary $Z_2$ spectrum can be detected. 
Here, we provide a qualitative analysis regarding the possibility of such GW observatories to detect this deviation. 
Two free parameters can affect the gravitational wave spectrum produced by the DW network: the wall tensions and the biases. 
As discussed in Sec.~\ref{sec:DW}, the DWs tension can be determined by $\beta$ up to an overall scale. 
In the following discussion, we fix the overall scale of the tensions by fixing the tension of SI and TI DW in the $Z_2$- and $Z_3$-preserving cases, respectively.
For the biases, we follow the ratio given in Tab.~\ref{tab:epsilons}, with the smallest bias as a free parameter.

To provide a more concrete discussion, we choose two benchmark values for the tensions of SI and TI DWs and the smallest biases, as shown in Tab.~\ref{tab:bps}. 
\begin{table}[]
 \centering
 \begin{tabular}{ccc}
 \hline\hline
 & $\sigma^{1/3}/\mathrm{TeV} $ & $V_{\rm bias}^{1/4}/{\rm TeV}$ \\ 
 BP1 & $ 3 \times 10^2$ & $10^{-3.75}$ \\
 BP2& $3 \times 10^5$& $10^1$\\
 \hline\hline
 \end{tabular}
 \caption{The values for the tension of SI and TI DWs and the smallest bias of the two benchmark points. }
 \label{tab:bps}
\end{table}
The estimated gravitational wave spectrum is shown in Fig.~\ref{fig:GW_sp}.
{The peak frequency of the GW signal is determined by a combination of the overall flavour symmetry breaking scale and the bias. The flavour predictions are independent of the breaking scale but are sensitive to the bias which acts as an explicit flavour symmetry breaking term that induces realistic flavour mixing. This topic was studied in detail in \cite{Gelmini:2020bqg}. As shown in the solid curves in Fig. 15 of our paper, one of the benchmark point, which is consistent with the
signal observed by PTA, can explain the flavour data as discussed in the previous reference with flavour symmetry breaking scale around $10^2$~TeV and right-handed neutrino mass of order 0.1~GeV, via Eq. (4.4) in \cite{Gelmini:2020bqg}. Interestingly, these scales are within reach of current intensity experiments \cite{Deppisch:2015qwa}, in particular, DUNE \cite{Ballett:2019bgd} and SHiP \cite{Agrawal:2021dbo,Antel:2023hkf}.} 
At the higher frequencies spanned by PTAs, the multi-peak structure shows up instead of the power-law structure, changing the slope of the spectrum after the peak. 
According to Ref.~\cite{NANOGrav:2023hvm}, the annihilation of the network happens at around 100 MeV.
For the other benchmark point, the peak falls within the frequency range that is detectable for next-generation space-based gravitational wave detectors, such as LISA and Taiji. The deviation from the ordinary $Z_2$ spectrum around the peak frequencies can also be observed.
Nevertheless, more quantitative discussion about the GW spectrum would require a numerical simulation of the DW dynamics and GW production, which will be addressed in subsequent works. {Moreover, it has been shown that DW generating a kiloHz frequency signal can also generate asteroid-mass primordial black holes which are a viable dark matter candidate \cite{Gouttenoire:2023gbn,Ferreira:2024eru}}

\section{Conclusion \label{sec:con}}
In this work, we have performed the first detailed analysis of the properties of domain walls (DWs) emerging from the spontaneous breaking of non-Abelian discrete symmetries. 
Choosing $S_4$ as an example, we have discussed the vacuum structure from a generic renormalisable potential of a triplet flavon $\phi$. 
Depending on the potential parameters, this group provides two types of vacua, six $Z_2$-preserving and eight $Z_3$-preserving vacua. 
From these vacua, we found five types of DWs, labelled as SI, SII and TI, TII, TIII, respectively. Here, SI and SII refer to two types DWs separating two $Z_2$-preserving, and TI, TII and TIII refer to three types of DWs separating two $Z_3$-preserving vacua. 

Given the different types of DWs, we explored their properties by computing the profiles, the tension and the thickness for each of them with respect to $\beta$ (where $\beta = g_2/g_1$), which is the only free parameter of $S_4$ after the redefinition of the scalar fields and the space coordinate. 
Noticing that the DW between some pairs of vacua can be unstable, we have discussed the stability of the various solutions, and we find the tension plays a crucial role. 
For $|\beta| \ll 1$ the difference between the tensions of the different DWs can be more than an order of magnitude. 
In that case, the DWs with higher tension will evolve into the ones with lower tension. 
The stable DW solutions are the SII type for the $Z_2$-preserving vacua and the TII type for the $Z_3$-preserving vacua.

Finally, we have estimated the gravitational wave signature for non-Abelian DWs and qualitatively compared it with the signal from a network of $Z_2$ DWs. 
We have found a multi-peak behaviour which is more enhanced with respect to what has been discussed in Ref.~\cite{Gelmini:2020bqg} because there are DW solutions with differing tensions which may be hierarchical.
When all the solutions are stable, i.e. $\beta > 2$ for the S-type DWs and $\beta < -1$ for the T-type DWs, the gravitational wave signals produced by the DWs with different tensions reach their peaks at various frequencies, which is a distinctive feature of DWs generated by non-Abelian discrete symmetries.

In summary, the DWs from non-Abelian discrete symmetry breaking have a richer structure than the $Z_2$ scenario in many aspects.
This could lead to interesting phenomena regarding the evolution of the DWs, such as solution instabilities that could affect the production of gravitational wave signals and may have other possible cosmological consequences. 
The study of DWs from non-Abelian discrete symmetries allows a precise connection between flavour physics and gravitational wave phenomenology and our work opens a new pathway to test flavour models.

\section*{Acknowledgement}

We want to thank Y. C. Wu for the helpful discussion. 
J.T. is supported by the STFC under Grant No.~ST/X003167/1.
This work was partially supported by European Union's Horizon 2020 Research,
Innovation Programme under Marie Sklodowska-Curie grant agreement HIDDeN European ITN project (H2020-MSCA-ITN-2019//860881-HIDDeN),
National Natural Science Foundation of China (NSFC) under Grant Nos. 12205064, 12347103, 
Zhejiang Provincial Natural Science Foundation of China under Grant No. LDQ24A050002,
STFC Consolidated Grant ST/T000775/1, the Spanish National Grant PID2022-137268NA-C55 and from the Generalitat Valenciana through the grant CIPROM/22/69, Guangdong Basic and Applied Basic Research Foundation No. 2025A1515011079.

\appendix

\section{Derivation of renormalisable potentials in $S_4$} \label{app:A}
The renormalisable potential $V(\phi)$ in Eq. \eqref{eq:Vvarphi} applies to $S_4$ in the Ma-Rajasekaran basis \cite{Ma:2001dn}. It is the general form of renormalisable potentials invariant in these discrete groups. Below we show how to derive it in $S_4$. 

Given two triplets $a=(a_1, a_2,a_3)^T$ and $b=(b_1, b_2, b_3)^T$ of $S_4$, 
In $S_4$, there are five irreps ${\bf 1}$, ${\bf 1}'$ ${\bf 2}$, ${\bf 3}$ and ${\bf 3}'$. The product of two ${\bf 3}^{(\prime)}$-plets follows the decomposition $\mathbf{3}\times\mathbf{3} = \mathbf{3}'\times\mathbf{3}' =\mathbf{1}+\mathbf{2}+\mathbf{3}+\mathbf{3}'$, with 
\begin{eqnarray} \label{eq:CG_S4}
(ab)_\mathbf{1}\;\, &=& a_1b_1 + a_2b_2 + a_3b_3 \,,\nonumber\\
(ab)_\mathbf{2}\, &=& (
 a_1b_1 + \omega a_2b_2 + \omega^2 a_3b_3, a_1b_1 + \omega^2 a_2b_2 + \omega a_3b_3 )^T \,,\nonumber\\
(ab)_{\mathbf{3}} &=& (a_2b_3+a_3b_2, a_3b_1+a_1b_3, a_1b_2+a_2b_1)^T \,,\nonumber\\
(ab)_{\mathbf{3}'} &=& (a_2b_3-a_3b_2, a_3b_1-a_1b_3, a_1b_2-a_2b_1)^T \,.
\end{eqnarray} 
The product of ${\bf 3}$ and ${\bf 3}'$ follows the decomposition $\mathbf{3}\times\mathbf{3}'=\mathbf{1}'+\mathbf{2}+\mathbf{3}'+\mathbf{3}$. Here no trivial singlet is given. Constructions of $(ab)_\mathbf{1}$ and $(ab)_\mathbf{2}$ follows those in $(ab)_\mathbf{1}$ and $(ab)_\mathbf{2}$ in Eq.~\eqref{eq:CG_S4}, respectively. Constructions of $(ab)_{\mathbf{3}}$ and $(ab)_{\mathbf{3}'}$ are reversed compared with those in Eq.~\eqref{eq:CG_S4}. 

If $\phi$ is arranged to be a triplet ${\bf 3}'$, the most general $S_4$-invariant potential is given by
\begin{eqnarray}
V(\phi)&=&- \frac{1}{2}\mu^2 (\phi \phi)_\mathbf{1} +\frac{1}{4} \left[f_1 \big( (\phi \phi)_\mathbf{1}\big)^2 + \frac12 f_2 \big( (\phi \phi)_{\mathbf{2}} (\phi \phi)_{\mathbf{2}} \big)_{\mathbf{1}} + f_3 \big( (\phi \phi)_{\mathbf{3}} (\phi \phi)_{\mathbf{3}} \big)_\mathbf{1} \right] \,,
\end{eqnarray} 
where $\mu^2$ and $f_{1,2,3}$ are real parameters. 
Then we fully recover Eq.~\eqref{eq:potential} with $g_1 = f_1+f_2$, $g_2 = 2 f_3- \frac23 f_2$. 
If $\phi$ is arranged to be the triplet ${\bf 3}$, where will also be a cubic term \big( $(\phi \phi)_{\mathbf{3}} \phi \big)_\mathbf{1} = 6 \phi_1 \phi_2 \phi_3$.

Including the cubic term actually gives the most general $A_4$-invariant renormalisable potential for a triplet. $A_4$ has four irreps, $\mathbf{1}$, $\mathbf{1}'$, $\mathbf{1}''$, and $\mathbf{3}$. 
Kronecker products of two triplet follows the decomposition $\mathbf{3}\times\mathbf{3}=\mathbf{1}+\mathbf{1}'+\mathbf{1}''+\mathbf{3}_S+\mathbf{3}_A$, where the subscripts $_S$ and $_A$ stand for the symmetric and anti-symmetric parts, respectively,
\begin{eqnarray}
(ab)_\mathbf{1}\;\, &=& a_1b_1 + a_2b_2 + a_3b_3 \,,\nonumber\\
(ab)_\mathbf{1'}\, &=& a_1b_1 + \omega a_2b_2 + \omega^2 a_3b_3 \,,\nonumber\\
(ab)_\mathbf{1''} &=& a_1b_1 + \omega^2 a_2b_2 + \omega a_3b_3 \,,\nonumber\\
(ab)_{\mathbf{3}_S} &=& (a_2b_3+a_3b_2, a_3b_1+a_1b_3, a_1b_2+a_2b_1)^T \,,\nonumber\\
(ab)_{\mathbf{3}_A} &=& (a_2b_3-a_3b_2, a_3b_1-a_1b_3, a_1b_2-a_2b_1)^T \,.
\end{eqnarray} 
Following the representation theory of $A_4$, one can construct the general $A_4$-invariant renormalisable potential as 
\begin{eqnarray}
V(\phi)&=&- \frac{1}{2}\mu^2 (\phi \phi)_\mathbf{1} +\frac{1}{4} \left[f_1 \big( (\phi \phi)_\mathbf{1}\big)^2 + f_2 (\phi \phi)_{\mathbf{1}'} (\phi \phi)_{\mathbf{1}''} + f_3 \big( (\phi \phi)_{\mathbf{3}_S} (\phi \phi)_{\mathbf{3}_S} \big)_\mathbf{1} \right] \nonumber\\ 
&&+ \frac{A}{6} \big( (\phi \phi)_{\mathbf{3}_S} \phi)_{\mathbf{3}_S} \big)_\mathbf{1} \,,
\label{eq:Vvarphi}
\end{eqnarray} 
where $A$ is real. This is almost the same except the last cubic term. In $A_4$, this term can be forbidden by introducing a parity $Z_2^P$ symmetry, $\phi \leftrightarrow -\phi$. 


\section{Domain wall properties at $\beta = 2$} \label{app:B}

In the special case at $\beta = 2$, i.e., $g_2 = 2 g_1$ in the potential. The domain wall can be analytically solved and we provide the of derivation below. 
We take the DW of $\mybox{v_1}\mybox{v_2}$ as an example. Since $\phi_3$ vanishes in the solution, it is enough for us to consider the effective potential of only $\phi_1$ and $\phi_2$, i.e.,
\begin{eqnarray}
 V(\phi_1, \phi_2) = - \frac{\mu^2}{2} (\phi_1^2 + \phi_2^2) + 
 \frac{g_1}{4} [(\phi_1^2 + \phi_2^2)^2 + 4 \phi_1^2 \phi_2^2]\,.
\end{eqnarray}
By taking the transformation $\phi_\pm = \frac{1}{\sqrt{2}}(\phi_1 \pm \phi_2)$, we obtain $V(\phi_1, \phi_2) = V(\phi_+) +V(\phi_-)$ with $V(\phi_\pm)$ given by
\begin{eqnarray}
 V(\phi_\pm) = - \frac{\mu^2}{2} \phi_\pm^2 + \frac{g_1}{2} \phi_\pm^4 \,.
\end{eqnarray}
In this parameterisation, $\phi_+$ and $\phi_-$ decouple in the potential. And the potential is invariant under two seperate $Z_2$ symmetries $Z_2^+:\phi_+ \leftrightarrow - \phi_+$ and $Z_2^-:\phi_- \leftrightarrow - \phi_-$. 
Boundary conditions for $\mybox{v_1}\mybox{v_2}$ are equivalent to 
\begin{eqnarray}
 \left.\phi_\pm \right|_{z\to \infty} = \frac{v}{\sqrt{2}} \,,\quad
 \left.\phi_{\pm} \right|_{z\to -\infty} = \pm \frac{v}{\sqrt{2}}
\end{eqnarray}
This suggests $\phi_+(z)$ takes the VEV $\frac{v}{\sqrt{2}}$ along the $z$-direction from $+\infty$ to $-\infty$ and $\phi_-(z)$ just follows the $Z_2$ domain wall solution $\phi_-(z) = \frac{v}{\sqrt{2}} \tanh (\frac{\mu z}{\sqrt{2}})$. 
The tension and thickness of the domain wall are obtained straightforwardly,
\begin{eqnarray}
 \sigma_{\text{SII}}(\beta = 2) = \mu \frac{v^2}{2} \frac{2\sqrt{2}}{3}\,, \quad
 \delta_{\text{SII}}(\beta = 2) = \frac{\sqrt{2}}{\mu} \,.
\end{eqnarray}
Following the parametrisation in the Sec.~\ref{sec:DWS}, we obtain the dimensionless tension and thickness:
\begin{eqnarray}
 \bar{\sigma}_{\text{SII}}(\beta = 2) = \frac{\sqrt{2}}{3} = \frac{1}{2} \bar{\sigma}_{\text{SI}} \,, \quad
 \bar{\delta}_{\text{SII}}(\beta = 2) = \sqrt{2} = \bar{\delta}_{\text{SI}} \,.
\end{eqnarray}
We have proved that at $\beta =2$, the energy stored in an SII DW is half that of a SI DW.

\section{Effect of the cubic term in the potential} \label{app:C}

By arranging the flavon as a ${\bf 3}$ of $S_4$ or considering the potential in $A_4$, we have to include a cubic term in the potential. Then the full renormalisable potential is written to be
\begin{eqnarray}
 V(\phi) = -\frac{\mu^2}{2} I_1 +\frac{g_1}{4} I_1^2 + \frac{g_2}{2} I_2 + A \phi_1 \phi_2 \phi_3 \,.
\end{eqnarray}
We discuss the vacuum structure and DW property induced by the last term. 

Including this term, the six $Z_2$-preserving vacua do not be modified, but the eight $Z_3$-preserving solutions $u_n$ are further split into to two groups, 
\begin{eqnarray}
u_{1,6,7,8} &=& 
\left\{\begin{pmatrix} 1 \\ 1 \\ 1 \end{pmatrix}, 
\begin{pmatrix} 1 \\ -1 \\ -1 \end{pmatrix},
\begin{pmatrix} -1 \\ 1 \\ -1 \end{pmatrix},
\begin{pmatrix} -1 \\ -1 \\ 1 \end{pmatrix}
\right\}u_-\,\label{eq:uminus}\\
u_{2,3,4,5} &=& 
\left\{\begin{pmatrix} -1 \\ 1 \\ 1 \end{pmatrix},
\begin{pmatrix} 1 \\ -1 \\ 1 \end{pmatrix},
\begin{pmatrix} 1 \\ 1 \\ -1 \end{pmatrix},
\begin{pmatrix} -1 \\ -1 \\ -1 \end{pmatrix}
\right\}u_+ \label{eq:uplus}\,,
\end{eqnarray}
where 
\begin{eqnarray}
&&u_{\mp}=\frac{\mu}{\sqrt{3g_1+2g_2}}\left( \sqrt{1+a^2}\mp a \right)\, \end{eqnarray}
and $a=\frac{A}{2\mu\sqrt{3g_1+2g_2}}$. The potential at these solutions are split into two values 
\begin{eqnarray}
V_{\mp}=-\frac{3\mu^4}{4 (3 g_1 + 2g_2)} \left(1+ \frac{2}{3} a^2 \mp \frac{2}{3} a \sqrt{a^2+1}\right) \left(\sqrt{a^2+1} \mp a \right)^2\,,
\end{eqnarray}
where the $\mp$ signs are for $u_{1,6,7,8}$ and $u_{2,3,4,5}$, respectively. 
In the case of negative $A$ (i.e., $a<0$), $V_-<V_+$, and thus $V_-$ is the global minimum. $u_{1,6,7,8}$ and $u_{2,3,4,5}$ are not equivalent any more and the former four solutions represent the true vacua, while the latter solutions become metastable false vacua. 
On the contrary, for positive $A$ (i.e., $a>0$), the vacua properties are flipped. 
In either cases, the flavon masses are modified to
\begin{eqnarray}
 && m_{1,\mp}^2 = 2\mu^2\left( 1+a^2 \mp a\sqrt{a^2+1} \right)\,,\nonumber\\
 && m_{2,\mp}^2=m_{3,\mp}^2 = - \frac{2 g_2}{3 g_1+2 g_2} \mu^2 \left[1 -2a\left(a \mp \sqrt{a^2+1} \right) (1+3 g_1/g_2)\right] \,.
\end{eqnarray}

Due to inclusion of the cubic term, the vacua structure become more complicated and we expect that DWs can have more complicated structures compared with those in the maintext. We will study this scenario in detail elsewhere. Below, we give brief comments on the influence of the cubic term on DWs. 

For S-type DWs, where two $Z_2$-preserving vacua have been fixed by boundary conditions, the coefficient $A$ of the cubic term does not contribute to the EOM of the scalar. 
Thus all S-type DWs are independent of the cubic term. 

For T-type DWs, the cubic term can contribute to their properties but at different levels for different types. 
For TIII DWs, the cubic term contributes to both the vacua expectation value and flavon EOM. 
Thus, it modifies the DWs just quantitatively.
Modifications to TI and TII DWs are more significant. 
A TI or TII DW separates two vacua in two different sets. 
We take $\mybox{u_1}\mybox{u_2}$ with $A<0$ as an example. 
From the discussion in the end of Sec.~\ref{sec:vac}, it is clear that the vacuum energy $V_-$ of the true vacua $u_1$ on one side is lower than the vacuum energy $V_+$ of the false vacua $u_2$ on the other side. 
Such a term, if it is not too large, can be treated as a bias. As discussed in section~\ref{sec:GW}, a bias makes DWs collapse at a certain temperature and generate gravitational waves.
Here, a non-zero cubic term can generate a pre-existing bias term to the vacua. If this term is larger than the bias from soft symmetry breaking, the TI and TII DWs, as well as half of the TIII DWs, may collapse at a higher temperature, namely, earlier than the other DWs. Its collapsing also generates GWs, but the signal is weak due to the large bias.

\bibliographystyle{JHEP}
\bibliography{Ref}

\providecommand{\href}[2]{#2}\begingroup\raggedright\begin{thebibliography}{10}

\bibitem{Kamiokande-II:1990wrs}
{\scshape Kamiokande-II} collaboration, {{Results from one thousand days of
  real time directional solar neutrino data}},
  \href{https://doi.org/10.1103/PhysRevLett.65.1297}{{Phys. Rev. Lett.}
  {\bfseries 65} (1990) 1297}.

\bibitem{Cleveland:1998nv}
B.T.~Cleveland, T.~Daily, R.~Davis, Jr., J.R.~Distel, K.~Lande, C.K.~Lee
  et~al., {{Measurement of the solar electron neutrino flux with the Homestake
  chlorine detector}}, \href{https://doi.org/10.1086/305343}{{Astrophys. J.}
  {\bfseries 496} (1998) 505}.

\bibitem{GALLEX:1998kcz}
{\scshape GALLEX} collaboration, {{GALLEX solar neutrino observations: Results
  for GALLEX IV}}, \href{https://doi.org/10.1016/S0370-2693(98)01579-2}{{Phys.
  Lett. B} {\bfseries 447} (1999) 127}.

\bibitem{SAGE:1999nng}
{\scshape SAGE} collaboration, {{Measurement of the solar neutrino capture rate
  with gallium metal}},
  \href{https://doi.org/10.1103/PhysRevC.60.055801}{{Phys. Rev. C} {\bfseries
  60} (1999) 055801} [\href{https://arxiv.org/abs/astro-ph/9907113}{{\ttfamily
  astro-ph/9907113}}].

\bibitem{Super-Kamiokande:2001ljr}
{\scshape Super-Kamiokande} collaboration, {{Solar B-8 and hep neutrino
  measurements from 1258 days of Super-Kamiokande data}},
  \href{https://doi.org/10.1103/PhysRevLett.86.5651}{{Phys. Rev. Lett.}
  {\bfseries 86} (2001) 5651}
  [\href{https://arxiv.org/abs/hep-ex/0103032}{{\ttfamily hep-ex/0103032}}].

\bibitem{Altarelli:2010gt}
G.~Altarelli and F.~Feruglio, {{Discrete Flavor Symmetries and Models of
  Neutrino Mixing}}, \href{https://doi.org/10.1103/RevModPhys.82.2701}{{Rev.
  Mod. Phys.} {\bfseries 82} (2010) 2701}
  [\href{https://arxiv.org/abs/1002.0211}{{\ttfamily 1002.0211}}].

\bibitem{King:2013eh}
S.F.~King and C.~Luhn, {{Neutrino Mass and Mixing with Discrete Symmetry}},
  \href{https://doi.org/10.1088/0034-4885/76/5/056201}{{Rept. Prog. Phys.}
  {\bfseries 76} (2013) 056201}
  [\href{https://arxiv.org/abs/1301.1340}{{\ttfamily 1301.1340}}].

\bibitem{King:2017guk}
S.F.~King, {{Unified Models of Neutrinos, Flavour and CP Violation}},
  \href{https://doi.org/10.1016/j.ppnp.2017.01.003}{{Prog. Part. Nucl. Phys.}
  {\bfseries 94} (2017) 217}
  [\href{https://arxiv.org/abs/1701.04413}{{\ttfamily 1701.04413}}].

\bibitem{Xing:2020ijf}
Z.-z.~Xing, {{Flavor structures of charged fermions and massive neutrinos}},
  \href{https://doi.org/10.1016/j.physrep.2020.02.001}{{Phys. Rept.} {\bfseries
  854} (2020) 1} [\href{https://arxiv.org/abs/1909.09610}{{\ttfamily
  1909.09610}}].

\bibitem{LIGOScientific:2016aoc}
{\scshape LIGO Scientific, Virgo} collaboration, {{Observation of Gravitational
  Waves from a Binary Black Hole Merger}},
  \href{https://doi.org/10.1103/PhysRevLett.116.061102}{{Phys. Rev. Lett.}
  {\bfseries 116} (2016) 061102}
  [\href{https://arxiv.org/abs/1602.03837}{{\ttfamily 1602.03837}}].

\bibitem{NANOGrav:2023gor}
{\scshape NANOGrav} collaboration, {{The NANOGrav 15 yr Data Set: Evidence for
  a Gravitational-wave Background}},
  \href{https://doi.org/10.3847/2041-8213/acdac6}{{Astrophys. J. Lett.}
  {\bfseries 951} (2023) L8}
  [\href{https://arxiv.org/abs/2306.16213}{{\ttfamily 2306.16213}}].

\bibitem{EPTA:2023fyk}
{\scshape EPTA, InPTA:} collaboration, {{The second data release from the
  European Pulsar Timing Array - III. Search for gravitational wave signals}},
  \href{https://doi.org/10.1051/0004-6361/202346844}{{Astron. Astrophys.}
  {\bfseries 678} (2023) A50}
  [\href{https://arxiv.org/abs/2306.16214}{{\ttfamily 2306.16214}}].

\bibitem{EPTA:2023xxk}
{\scshape EPTA} collaboration, {{The second data release from the European
  Pulsar Timing Array: V. Implications for massive black holes, dark matter and
  the early Universe}},  \href{https://arxiv.org/abs/2306.16227}{{\ttfamily
  2306.16227}}.

\bibitem{Reardon:2023gzh}
D.J.~Reardon et~al., {{Search for an Isotropic Gravitational-wave Background
  with the Parkes Pulsar Timing Array}},
  \href{https://doi.org/10.3847/2041-8213/acdd02}{{Astrophys. J. Lett.}
  {\bfseries 951} (2023) L6}
  [\href{https://arxiv.org/abs/2306.16215}{{\ttfamily 2306.16215}}].

\bibitem{Xu:2023wog}
H.~Xu et~al., {{Searching for the Nano-Hertz Stochastic Gravitational Wave
  Background with the Chinese Pulsar Timing Array Data Release I}},
  \href{https://doi.org/10.1088/1674-4527/acdfa5}{{Res. Astron. Astrophys.}
  {\bfseries 23} (2023) 075024}
  [\href{https://arxiv.org/abs/2306.16216}{{\ttfamily 2306.16216}}].

\bibitem{Madge:2023dxc}
E.~Madge, E.~Morgante, C.~Puchades-Ib\'a\~nez, N.~Ramberg, W.~Ratzinger,
  S.~Schenk et~al., {{Primordial gravitational waves in the nano-Hertz regime
  and PTA data \textemdash{} towards solving the GW inverse problem}},
  \href{https://doi.org/10.1007/JHEP10(2023)171}{{JHEP} {\bfseries 10} (2023)
  171} [\href{https://arxiv.org/abs/2306.14856}{{\ttfamily 2306.14856}}].

\bibitem{Vilenkin:1984ib}
A.~Vilenkin, {{Cosmic Strings and Domain Walls}},
  \href{https://doi.org/10.1016/0370-1573(85)90033-X}{{Phys. Rept.} {\bfseries
  121} (1985) 263}.

\bibitem{Kibble:1976sj}
T.W.B.~Kibble, {{Topology of Cosmic Domains and Strings}},
  \href{https://doi.org/10.1088/0305-4470/9/8/029}{{J. Phys. A} {\bfseries 9}
  (1976) 1387}.

\bibitem{Preskill:1984gd}
J.~Preskill, {{MAGNETIC MONOPOLES}},
  \href{https://doi.org/10.1146/annurev.ns.34.120184.002333}{{Ann. Rev. Nucl.
  Part. Sci.} {\bfseries 34} (1984) 461}.

\bibitem{Auclair:2019wcv}
P.~Auclair et~al., {{Probing the gravitational wave background from cosmic
  strings with LISA}},
  \href{https://doi.org/10.1088/1475-7516/2020/04/034}{{JCAP} {\bfseries 04}
  (2020) 034} [\href{https://arxiv.org/abs/1909.00819}{{\ttfamily
  1909.00819}}].

\bibitem{Gouttenoire:2019kij}
Y.~Gouttenoire, G.~Servant and P.~Simakachorn, {{Beyond the Standard Models
  with Cosmic Strings}},
  \href{https://doi.org/10.1088/1475-7516/2020/07/032}{{JCAP} {\bfseries 07}
  (2020) 032} [\href{https://arxiv.org/abs/1912.02569}{{\ttfamily
  1912.02569}}].

\bibitem{Hiramatsu:2010yz}
T.~Hiramatsu, M.~Kawasaki and K.~Saikawa, {{Gravitational Waves from Collapsing
  Domain Walls}}, \href{https://doi.org/10.1088/1475-7516/2010/05/032}{{JCAP}
  {\bfseries 05} (2010) 032} [\href{https://arxiv.org/abs/1002.1555}{{\ttfamily
  1002.1555}}].

\bibitem{Kawasaki:2011vv}
M.~Kawasaki and K.~Saikawa, {{Study of gravitational radiation from cosmic
  domain walls}}, \href{https://doi.org/10.1088/1475-7516/2011/09/008}{{JCAP}
  {\bfseries 09} (2011) 008} [\href{https://arxiv.org/abs/1102.5628}{{\ttfamily
  1102.5628}}].

\bibitem{Hiramatsu:2013qaa}
T.~Hiramatsu, M.~Kawasaki and K.~Saikawa, {{On the estimation of gravitational
  wave spectrum from cosmic domain walls}},
  \href{https://doi.org/10.1088/1475-7516/2014/02/031}{{JCAP} {\bfseries 02}
  (2014) 031} [\href{https://arxiv.org/abs/1309.5001}{{\ttfamily 1309.5001}}].

\bibitem{Saikawa:2017hiv}
K.~Saikawa, {{A review of gravitational waves from cosmic domain walls}},
  \href{https://doi.org/10.3390/universe3020040}{{Universe} {\bfseries 3}
  (2017) 40} [\href{https://arxiv.org/abs/1703.02576}{{\ttfamily 1703.02576}}].

\bibitem{Buchmuller:2021mbb}
W.~Buchmuller, V.~Domcke and K.~Schmitz, {{Stochastic gravitational-wave
  background from metastable cosmic strings}},
  \href{https://doi.org/10.1088/1475-7516/2021/12/006}{{JCAP} {\bfseries 12}
  (2021) 006} [\href{https://arxiv.org/abs/2107.04578}{{\ttfamily
  2107.04578}}].

\bibitem{Dunsky:2021tih}
D.I.~Dunsky, A.~Ghoshal, H.~Murayama, Y.~Sakakihara and G.~White, {{GUTs,
  hybrid topological defects, and gravitational waves}},
  \href{https://doi.org/10.1103/PhysRevD.106.075030}{{Phys. Rev. D} {\bfseries
  106} (2022) 075030} [\href{https://arxiv.org/abs/2111.08750}{{\ttfamily
  2111.08750}}].

\bibitem{King:2020hyd}
S.F.~King, S.~Pascoli, J.~Turner and Y.-L.~Zhou, {{Gravitational Waves and
  Proton Decay: Complementary Windows into Grand Unified Theories}},
  \href{https://doi.org/10.1103/PhysRevLett.126.021802}{{Phys. Rev. Lett.}
  {\bfseries 126} (2021) 021802}
  [\href{https://arxiv.org/abs/2005.13549}{{\ttfamily 2005.13549}}].

\bibitem{King:2021gmj}
S.F.~King, S.~Pascoli, J.~Turner and Y.-L.~Zhou, {{Confronting SO(10) GUTs with
  proton decay and gravitational waves}},
  \href{https://doi.org/10.1007/JHEP10(2021)225}{{JHEP} {\bfseries 10} (2021)
  225} [\href{https://arxiv.org/abs/2106.15634}{{\ttfamily 2106.15634}}].

\bibitem{Fu:2022lrn}
B.~Fu, S.F.~King, L.~Marsili, S.~Pascoli, J.~Turner and Y.-L.~Zhou, {{A
  predictive and testable unified theory of fermion masses, mixing and
  leptogenesis}}, \href{https://doi.org/10.1007/JHEP11(2022)072}{{JHEP}
  {\bfseries 11} (2022) 072}
  [\href{https://arxiv.org/abs/2209.00021}{{\ttfamily 2209.00021}}].

\bibitem{Fu:2023mdu}
B.~Fu, S.F.~King, L.~Marsili, S.~Pascoli, J.~Turner and Y.-L.~Zhou, {{Testing
  realistic SO(10) SUSY GUTs with proton decay and gravitational waves}},
  \href{https://doi.org/10.1103/PhysRevD.109.055025}{{Phys. Rev. D} {\bfseries
  109} (2024) 055025} [\href{https://arxiv.org/abs/2308.05799}{{\ttfamily
  2308.05799}}].

\bibitem{NANOGrav:2023hvm}
{\scshape NANOGrav} collaboration, {{The NANOGrav 15 yr Data Set: Search for
  Signals from New Physics}},
  \href{https://doi.org/10.3847/2041-8213/acdc91}{{Astrophys. J. Lett.}
  {\bfseries 951} (2023) L11}
  [\href{https://arxiv.org/abs/2306.16219}{{\ttfamily 2306.16219}}].

\bibitem{Kitajima:2023cek}
N.~Kitajima, J.~Lee, K.~Murai, F.~Takahashi and W.~Yin, {{Gravitational waves
  from domain wall collapse, and application to nanohertz signals with
  QCD-coupled axions}},
  \href{https://doi.org/10.1016/j.physletb.2024.138586}{{Phys. Lett. B}
  {\bfseries 851} (2024) 138586}
  [\href{https://arxiv.org/abs/2306.17146}{{\ttfamily 2306.17146}}].

\bibitem{Bai:2023cqj}
Y.~Bai, T.-K.~Chen and M.~Korwar, {{QCD-collapsed domain walls: QCD phase
  transition and gravitational wave spectroscopy}},
  \href{https://doi.org/10.1007/JHEP12(2023)194}{{JHEP} {\bfseries 12} (2023)
  194} [\href{https://arxiv.org/abs/2306.17160}{{\ttfamily 2306.17160}}].

\bibitem{Babichev:2023pbf}
E.~Babichev, D.~Gorbunov, S.~Ramazanov, R.~Samanta and A.~Vikman, {{NANOGrav
  spectral index \ensuremath{\gamma}=3 from melting domain walls}},
  \href{https://doi.org/10.1103/PhysRevD.108.123529}{{Phys. Rev. D} {\bfseries
  108} (2023) 123529} [\href{https://arxiv.org/abs/2307.04582}{{\ttfamily
  2307.04582}}].

\bibitem{Zhang:2023nrs}
Z.~Zhang, C.~Cai, Y.-H.~Su, S.~Wang, Z.-H.~Yu and H.-H.~Zhang, {{Nano-Hertz
  gravitational waves from collapsing domain walls associated with freeze-in
  dark matter in light of pulsar timing array observations}},
  \href{https://doi.org/10.1103/PhysRevD.108.095037}{{Phys. Rev. D} {\bfseries
  108} (2023) 095037} [\href{https://arxiv.org/abs/2307.11495}{{\ttfamily
  2307.11495}}].

\bibitem{Blasi:2023sej}
S.~Blasi, A.~Mariotti, A.~Rase and A.~Sevrin, {{Axionic domain walls at Pulsar
  Timing Arrays: QCD bias and particle friction}},
  \href{https://doi.org/10.1007/JHEP11(2023)169}{{JHEP} {\bfseries 11} (2023)
  169} [\href{https://arxiv.org/abs/2306.17830}{{\ttfamily 2306.17830}}].

\bibitem{Gouttenoire:2023ftk}
Y.~Gouttenoire and E.~Vitagliano, {{Domain wall interpretation of the PTA
  signal confronting black hole overproduction}},
  \href{https://doi.org/10.1103/PhysRevD.110.L061306}{{Phys. Rev. D} {\bfseries
  110} (2024) L061306} [\href{https://arxiv.org/abs/2306.17841}{{\ttfamily
  2306.17841}}].

\bibitem{Ferreira:2024eru}
R.Z.~Ferreira, A.~Notari, O.~Pujol\`as and F.~Rompineve, {{Collapsing domain
  wall networks: impact on pulsar timing arrays and primordial black holes}},
  \href{https://doi.org/10.1088/1475-7516/2024/06/020}{{JCAP} {\bfseries 06}
  (2024) 020} [\href{https://arxiv.org/abs/2401.14331}{{\ttfamily
  2401.14331}}].

\bibitem{Wu:2022tpe}
Y.~Wu, K.-P.~Xie and Y.-L.~Zhou, {{Classification of Abelian domain walls}},
  \href{https://doi.org/10.1103/PhysRevD.106.075019}{{Phys. Rev. D} {\bfseries
  106} (2022) 075019} [\href{https://arxiv.org/abs/2205.11529}{{\ttfamily
  2205.11529}}].

\bibitem{Wu:2022stu}
Y.~Wu, K.-P.~Xie and Y.-L.~Zhou, {{Collapsing domain walls beyond Z2}},
  \href{https://doi.org/10.1103/PhysRevD.105.095013}{{Phys. Rev. D} {\bfseries
  105} (2022) 095013} [\href{https://arxiv.org/abs/2204.04374}{{\ttfamily
  2204.04374}}].

\bibitem{Gelmini:2020bqg}
G.B.~Gelmini, S.~Pascoli, E.~Vitagliano and Y.-L.~Zhou, {{Gravitational wave
  signatures from discrete flavor symmetries}},
  \href{https://doi.org/10.1088/1475-7516/2021/02/032}{{JCAP} {\bfseries 02}
  (2021) 032} [\href{https://arxiv.org/abs/2009.01903}{{\ttfamily
  2009.01903}}].

\bibitem{Jueid:2023cgp}
A.~Jueid, M.A.~Loualidi, S.~Nasri and M.A.~Ouahid, {{Cosmological domain walls
  from the breaking of S4 flavor symmetry}},
  \href{https://doi.org/10.1103/PhysRevD.109.055048}{{Phys. Rev. D} {\bfseries
  109} (2024) 055048} [\href{https://arxiv.org/abs/2312.04388}{{\ttfamily
  2312.04388}}].

\bibitem{Yang:2024bys}
Y.~Yang and I.P.~Ivanov, {{Charge-breaking domain walls separating neutral
  vacua in multi-Higgs models}},
  \href{https://doi.org/10.1103/PhysRevD.110.015001}{{Phys. Rev. D} {\bfseries
  110} (2024) 015001} [\href{https://arxiv.org/abs/2401.03264}{{\ttfamily
  2401.03264}}].

\bibitem{Hagedorn:2012ut}
C.~Hagedorn, S.F.~King and C.~Luhn, {{SUSY S$_{4} \times$ SU(5) revisited}},
  \href{https://doi.org/10.1016/j.physletb.2012.09.026}{{Phys. Lett. B}
  {\bfseries 717} (2012) 207}
  [\href{https://arxiv.org/abs/1205.3114}{{\ttfamily 1205.3114}}].

\bibitem{Ding:2013eca}
G.-J.~Ding and Y.-L.~Zhou, {{Dirac Neutrinos with $S_4$ Flavor Symmetry in
  Warped Extra Dimensions}},
  \href{https://doi.org/10.1016/j.nuclphysb.2013.08.011}{{Nucl. Phys. B}
  {\bfseries 876} (2013) 418}
  [\href{https://arxiv.org/abs/1304.2645}{{\ttfamily 1304.2645}}].

\bibitem{Luhn:2013lkn}
C.~Luhn, {{Trimaximal TM$_{1}$ neutrino mixing in S$_{4}$ with spontaneous CP
  violation}}, \href{https://doi.org/10.1016/j.nuclphysb.2013.07.003}{{Nucl.
  Phys. B} {\bfseries 875} (2013) 80}
  [\href{https://arxiv.org/abs/1306.2358}{{\ttfamily 1306.2358}}].

\bibitem{King:2016yvg}
S.F.~King and C.~Luhn, {{Littlest Seesaw model from S$_{4} \times$ U(1)}},
  \href{https://doi.org/10.1007/JHEP09(2016)023}{{JHEP} {\bfseries 09} (2016)
  023} [\href{https://arxiv.org/abs/1607.05276}{{\ttfamily 1607.05276}}].

\bibitem{Ma:2001dn}
E.~Ma and G.~Rajasekaran, {{Softly broken A(4) symmetry for nearly degenerate
  neutrino masses}}, \href{https://doi.org/10.1103/PhysRevD.64.113012}{{Phys.
  Rev. D} {\bfseries 64} (2001) 113012}
  [\href{https://arxiv.org/abs/hep-ph/0106291}{{\ttfamily hep-ph/0106291}}].

\bibitem{Altarelli:2005yx}
G.~Altarelli and F.~Feruglio, {{Tri-bimaximal neutrino mixing, A(4) and the
  modular symmetry}},
  \href{https://doi.org/10.1016/j.nuclphysb.2006.02.015}{{Nucl. Phys. B}
  {\bfseries 741} (2006) 215}
  [\href{https://arxiv.org/abs/hep-ph/0512103}{{\ttfamily hep-ph/0512103}}].

\bibitem{Wainwright:2011kj}
C.L.~Wainwright, {{CosmoTransitions: Computing Cosmological Phase Transition
  Temperatures and Bubble Profiles with Multiple Fields}},
  \href{https://doi.org/10.1016/j.cpc.2012.04.004}{{Comput. Phys. Commun.}
  {\bfseries 183} (2012) 2006}
  [\href{https://arxiv.org/abs/1109.4189}{{\ttfamily 1109.4189}}].

\bibitem{Preskill:1991kd}
J.~Preskill, S.P.~Trivedi, F.~Wilczek and M.B.~Wise, {{Cosmology and broken
  discrete symmetry}},
  \href{https://doi.org/10.1016/0550-3213(91)90241-O}{{Nucl. Phys. B}
  {\bfseries 363} (1991) 207}.

\bibitem{Riva:2010jm}
F.~Riva, {{Low-Scale Leptogenesis and the Domain Wall Problem in Models with
  Discrete Flavor Symmetries}},
  \href{https://doi.org/10.1016/j.physletb.2010.05.073}{{Phys. Lett. B}
  {\bfseries 690} (2010) 443}
  [\href{https://arxiv.org/abs/1004.1177}{{\ttfamily 1004.1177}}].

\bibitem{Ouahid:2018gpg}
M.A.~Ouahid, M.A.~Loualidi, R.A.~Laamara and E.H.~Saidi, {{Neutrino
  phenomenology in the flavored NMSSM without domain wall problems}},
  \href{https://doi.org/10.1103/PhysRevD.102.115023}{{Phys. Rev. D} {\bfseries
  102} (2020) 115023} [\href{https://arxiv.org/abs/1810.10753}{{\ttfamily
  1810.10753}}].

\bibitem{Deppisch:2015qwa}
F.F.~Deppisch, P.S.~Bhupal~Dev and A.~Pilaftsis, {{Neutrinos and Collider
  Physics}}, \href{https://doi.org/10.1088/1367-2630/17/7/075019}{{New J.
  Phys.} {\bfseries 17} (2015) 075019}
  [\href{https://arxiv.org/abs/1502.06541}{{\ttfamily 1502.06541}}].

\bibitem{Ballett:2019bgd}
P.~Ballett, T.~Boschi and S.~Pascoli, {{Heavy Neutral Leptons from low-scale
  seesaws at the DUNE Near Detector}},
  \href{https://doi.org/10.1007/JHEP03(2020)111}{{JHEP} {\bfseries 03} (2020)
  111} [\href{https://arxiv.org/abs/1905.00284}{{\ttfamily 1905.00284}}].

\bibitem{Agrawal:2021dbo}
P.~Agrawal et~al., {{Feebly-interacting particles: FIPs 2020 workshop report}},
  \href{https://doi.org/10.1140/epjc/s10052-021-09703-7}{{Eur. Phys. J. C}
  {\bfseries 81} (2021) 1015}
  [\href{https://arxiv.org/abs/2102.12143}{{\ttfamily 2102.12143}}].

\bibitem{Antel:2023hkf}
C.~Antel et~al., {{Feebly-interacting particles: FIPs 2022 Workshop Report}},
  \href{https://doi.org/10.1140/epjc/s10052-023-12168-5}{{Eur. Phys. J. C}
  {\bfseries 83} (2023) 1122}
  [\href{https://arxiv.org/abs/2305.01715}{{\ttfamily 2305.01715}}].

\bibitem{Gouttenoire:2023gbn}
Y.~Gouttenoire and E.~Vitagliano, {{Primordial black holes and wormholes from
  domain wall networks}},
  \href{https://doi.org/10.1103/PhysRevD.109.123507}{{Phys. Rev. D} {\bfseries
  109} (2024) 123507} [\href{https://arxiv.org/abs/2311.07670}{{\ttfamily
  2311.07670}}].

\end{thebibliography}\endgroup

\end{document}